\documentclass{aa}
\pdfoutput=1
\usepackage{float}
\usepackage{graphicx,natbib,txfonts,color}
\usepackage{minipage-marginpar}
\usepackage{amssymb, amsmath}
\usepackage[separate-uncertainty=true]{siunitx}
\usepackage{multicol}
\usepackage{placeins}
\usepackage[colorlinks=true,allcolors=blue]{hyperref}

\bibpunct{(}{)}{;}{a}{}{,}

\newcommand{\rjup}{\mathrm{R}_{\text{Jup}}}

\newcommand{\rsun}{\mathrm{R}_{\odot}}

\newcommand{\mplanet}{\mathrm{M}_{\text{pl}}}
\newcommand{\rplanet}{\mathrm{R}_{\text{pl}}}

\newcommand{\rstar}{\mathrm{R}_{\star}}
\newcommand{\AU}{\text{AU}}
\newcommand{\tcol}{\mathrm{t}_{\mathrm{col}}}

\newcommand{\Kzz}{\mathrm{K}_{zz}}
\newcommand{\vwind}{\mathrm{u}_\text{wind}}
\newcommand{\Pwind}{\mathrm{P}_\text{wind}}

\newcommand{\Teff}{\mathrm{T}_\text{eff}}

\newcommand{\tauzonal}{\tau_\text{zonal}}

\newcommand{\cmspersquared}{\, \text{cm}/\text{s}^2}

\newcommand{\kms}{\, \text{km/s}}

\newcommand{\degrees}{^\circ}

\newcommand{\um}{\, \mu\text{m}}
\newcommand{\mbar}{\, \text{mbar}}
\newcommand{\mubar}{\, \mu\text{bar}}

\newcommand{\ammonia}{\mathrm{NH}_3}
\newcommand{\NHthree}{\mathrm{NH}_3}
\newcommand{\NHtwo}{\mathrm{NH}_2}
\newcommand{\CO}{\text{CO}}
\newcommand{\OH}{\mathrm{OH}}
\newcommand{\COtwo}{\mathrm{CO}_2}
\newcommand{\CHfour}{\mathrm{CH}_4}
\newcommand{\methane}{\mathrm{CH}_4}
\newcommand{\HCN}{\mathrm{HCN}}
\newcommand{\acetylene}{\mathrm{C}_2\mathrm{H}_2}
\newcommand{\water}{\mathrm{H}_2\mathrm{O}}
\newcommand{\Htwo}{\mathrm{H}_2}

\newcommand{\He}{\mathrm{He}}
\newcommand{\Ntwo}{\mathrm{N}_2}

\graphicspath{{./plots/}}

\begin{document}

%\thesaurus{06(08.01.1; 08.03.1; 08.05.3; 08.16.4)}

\title{Impact of stellar flares on the chemical composition and transmission spectra of gaseous exoplanets orbiting M dwarfs}
%Exoplanet atmospheres, tidally locked, 2D, chemistry, stellar flares
%\subtitle{The impact of flaring stellar irradiation}

\author{T. Konings  \inst{1},
	R. Baeyens \inst{1},
\and L. Decin \inst{1}}

%\offprints{T. Konings, \email{thomas.konings@kuleuven.be}}

\institute{${}^1$Institute of Astronomy, KU Leuven, Celestijnenlaan 200D,
3001 Leuven, Belgium }
%\date{Accepted 2 August 2022}
\date{Received 28 February 2022 / Accepted 2 August 2022}

\authorrunning{T. Konings et al.}
\titlerunning{Impact of stellar flares on gaseous exoplanets}

%*****************************************************************************
%                      ABSTRACT
%*****************************************************************************
\abstract
%context
{Stellar flares of active M dwarfs can affect the atmospheric composition of close-orbiting gas giants, and can result in time-dependent transmission spectra.
}
%aims
{We aim to examine the impact of a variety of flares, differing in energy, duration, and occurrence frequency, on the composition and transmission spectra of close-orbiting, tidally locked gaseous planets with climates dominated by equatorial superrotation. 
}
%Methods
{
We used a series of pseudo-2D photo- and thermochemical kinetics models, which take advection by the equatorial jet stream into account, 
to simulate the neutral molecular composition of a gaseous planet ($\Teff =800$ K) that orbits a M dwarf
during artificially constructed flare events.
We then computed transmission spectra for the evening and morning limb.
}
%Results
{
We find that the upper regions (i.e. below $10 \mubar$) of the dayside and evening limb are heavily depleted in $\CHfour$ and $\ammonia$ up to 
several days 
after a flare event with a total radiative energy of $\num{2e33}$ erg. 
Molar fractions of $\acetylene$ and $\HCN$ are enhanced up to a factor three on the nightside and morning limb 
after day-to-nightside advection of photodissociated $\CHfour$ and $\NHthree$.
Methane depletion reduces transit depths by 100-300 parts per million (ppm) on the evening limb 
and $\acetylene$ production increases the 14 $\um$ feature up to 350 ppm on the morning limb.
We find that repeated flaring drives the atmosphere to a composition that differs from its pre-flare distribution and that this translates to a permanent modification of the transmission spectrum.
}
%Conclusion
{
We show that single high-energy flares can affect the atmospheres of close-orbiting gas giants up to several days after the flare event, during which their transmission spectra are altered by several hundred ppm.
Repeated flaring has important implications for future retrieval analyses of exoplanets around active stars, as the atmospheric composition and resulting spectral signatures substantially differ from models that do not include flaring.
}

%\keywords{interstellar medium: jets and outflows --
%	interstellar medium: molecules -- stars: pre-main-sequence}}

\maketitle

%******************************************************************
%                    INTRODUCTION
%******************************************************************

\section{Introduction}\label{sec:intro}
%M dwarfs are the most abundant type of stars in the Galaxy \citep{Bochanski2010-TheLuminosityandMass} and are therefore ideal targets in the search for exoplanets.
In the search for new exoplanets, M dwarfs are ideal targets due to their high abundance in the Galaxy \citep{Bochanski2010-TheLuminosityandMass}.
%Together with their small masses and radii, M dwarfs also allow for atmospheric characterisation by high signal-to-noise transmission spectroscopy.
%They have small radii, resulting in good SNR when taking transmission spectra.
%They are therefore also favoured candidates in the search for smaller, rocky and potentially habitable planets.
However, M-type stars are prone to high levels of stellar activity \citep{Walkowicz2011-WhitelightFlares, Loyd2016-MUSCLESTreasurySruveyIII, Loyd2018-HAZMETIV} that can impact the radial velocity and/or transit signal of such systems through phenomena such as flaring \citep{Tofflemire2012-TheImplicationsofMDwarf}, star spots, plages and faculae \citep{Boisse2011-Disentangling,Llama2015-TransitingtheSun, Cauley2018-TheEffectsofstellar, Roettenbacher2022-EXPRESIII, Bruno2022-Hidinginplainsight}, and other activity-induced variability \citep{Dumusque2018-Measuringpreciseradial, Rackham2019-ConstrainingStellarPhotospheres, Bellotti2021-Mitigatingstellaractivityjitter, CollierCameron2021-Separatingpl}.
Aside from the observational implications, the planet's physical and chemical state can be altered by stellar activity as well due to, for example, coronal mass ejections (CMEs) and stellar particle events (SPEs) \citep{Yamashiki2019-ImpactofStellar, Atri2017-Modellingstellarproton, Atri2020-StellarEnergeticParticles, Segura2010-TheEffectofaStrong}, winds \citep{Vidotto2015-ontheenvironment, Vidotto2020-Stellarwindeffects, Chebly2021-Destinationexoplane, Colombo2021-HotJupitersaccreting}, and stellar flares \citep{Segura2010-TheEffectofaStrong, Venot2016-Influenceof, Chadney2017-Effectofstellarflares, Tilley2019-ModelingRepeatedMDwarf, Chen2021-Persistenceofflaredriven, Louca2022+Theimpactof}, the latter being sudden releases of radiative energy triggered by magnetic reconnection \citep{Benz2010-PhysicalProcessesinMagnetically}.
Stellar flares result in a temporary increase in incident flux on the planet's atmosphere, which in turn increases the photochemical reaction rates that can change the chemical composition.
Photochemistry, and photolysis in particular, is a key driver of chemical disequilibrium in the atmospheres of close-orbiting, gaseous planets.
Photolysis does not only deplete the upper layers from species such as $\CHfour$ and $\NHthree$, but it can enrich the middle regions with haze precursors such as $\HCN$ and $\acetylene$ as well, particularly on cooler planets \citep{Moses2011-Disequilibrium, Venot2012-Achemicalmodel,Zahnle2014-MethaneCarbonMonoxid,  Agundez2014-Pseudo2Dchemicalmodel, Moses2014-Chemicalkinetics, Miguel2014-ExoploringAtmospheresof, Rimmer2016-AChemicalKinetics,  Drummond2016-effectsofconsistent, Hobbs2019-Achemicalkineticscode,  Shulyak2020-Stellarimpact, Barth2021--MOVESIV, Baeyens2022-GridIIphotochemistry}.
Stellar flares thus have the potential to alter the chemical composition and, subsequently, alter the atmosphere's signature in transmission spectra. %
%particluarly on cooler planets because of long chemical timescales and high abudances of photochemically active species 
%Photochemistry that alters the middle/deep atmosphere also impacts observables such as transit depths \citep{Baeyens2022-GridIIphotochemistry}.
%
%
%Although the effect of flaring M dwarfs on exoplanet atmospheres has mostly been studied for rocky exoplanets, flaring can potentially impact close-orbiting gaseous planets by temporary enhancing photochemical activity.
%This can then result in a different transmission spectra compared to the pre-flare signal, which can result in \textit{wrong retrieved steady-state abundances}.
%Space missions such as JWST, PLATO and ARIEL will reach 50 to 100 ppm precision level.
%It is important to understand processes that affect the transmission spectra to this level in order to prevent biased retrievals that result in wrong steady-state abundances.

%\textcolor{red}{Should we mention somewhere that giant planets around low-mass stars are rare?
%\citep{Mulders2018-PlanetPopulations, Howard2012-PlanetOccurrencewithin, Johnson2010-GiantPlanetOccurrence, Wright2012-FrequencyHotJupiters, Cumming2008-KeckPlanetSearch} }

%\newline
%\newline

%\noindent
Several studies have been conducted that have used photo- and thermochemical kinetics models to study the effect of flares on exoplanetary atmospheres.
\citet{Segura2010-TheEffectofaStrong} tracked the water and ozone abundances of an Earth-like planetary atmosphere during a flare and SPE of its hypothetical M-type host star to address the consequences on the planet's habitability.
The authors constructed a time series from observations of the great flare event (GFE) of AD Leo, observed by \citet{Hawley1991-TheGreatFlare} on April 12, 1985.
They found a strong depletion of water vapour at high altitudes and fluctuation in the ozone column depth of less than 1\%, and thus concluded that a flare is insufficient to be a biological hazard.
An accompanying proton event poses more danger as nitrogen oxides are formed at the cost of ozone, leaving the planetary surface unshielded from incident UV radiation.
\citet{Tilley2019-ModelingRepeatedMDwarf} performed a follow-up study considering repeated flaring. 
They found that multiple SPEs can reduce the initial ozone column depth of an Earth-like planet to 6\% of its initial value in 10 years time, while strictly radiative flaring overall has a limited impact.
More recently, \citet{Chen2021-Persistenceofflaredriven} performed extensive three-dimensional coupled chemistry-climate model (CCM) simulations for rocky planets orbiting flaring G-, K-, and M-type stars.
They found that recurring flares of K and M dwarfs drive the atmospheric composition into new equilibria that deviate from the pre-flare values (e.g. for various nitrogen and hydrogen oxides, ozone, and water vapour), but they find overall variations to the transmission spectra of less than 10 parts per million (ppm).

The impact of stellar flares has also been examined for H/He-dominated atmospheres of gas giants.
\citet{Venot2016-Influenceof} studied the consequences of AD Leo's GFE for a sub-Neptune ($\Teff=1303$ K) and super-Earth ($\Teff=412$ K).
They found that prominent species in the atmosphere (such as $\water$, $\OH$, $\COtwo$, $\NHtwo$, $\mathrm{NO}$, and $\NHthree$) are affected.
During the flare event, the transit depths of the super-Earth vary with less than 15 ppm, while the sub-Neptune experiences variations up to 150 ppm, situated in the $\CO$/$\COtwo$ bumps at $4.6$ and $14 \um$.
More interestingly, their post-flare compositions converge to a new steady-state after \num{e12} seconds ($\sim$\num{30 000} years) which results in transmission spectra that are different from their pre-flare state by up to 40 ppm and 500 ppm for the super-Earth and sub-Neptune, respectively.
Additionally, \citet{Venot2016-Influenceof} explored the possibility of a repeated flaring every five hours and found that this leads to variations up to 1200 ppm after only one day of such periodic flaring.
%\textit{Finally,} \citet{Chadney2017-Effectofstellarflares} \textit{studied the effects of stellar flares on HD 209458b and HD 189733b, focussing on the induced thermal escape.
%They find that the neutral atmosphere of both planets remains relativity unaffected, while an extreme stellar proton event could result in a non-negligible mass loss.}
%\newline
%\newline
Finally, taking into account the frequency of the flare's occurrence as a function of its radiative energy 
%(the flare frequency distribution, FFD) 
could lead to accumulative changes in the chemical composition for both $\Htwo-$ and $\Ntwo$-dominated atmospheres, although the transmission spectra remain relatively unaffected (< 12 ppm) \citep{Louca2022+Theimpactof}. 
%\{Finally, \citet{Louca2022+Theimpactof} showed that the incorporation of a flare frequency distribution or FFD (i.e. the frequency of the flare's occurrence versus it's radiative energy) could lead to accumulative changes in the chemical composition for both $\Htwo-$ and $\Ntwo$-dominated atmospheres, although the transmission spectra remain relatively unaffected (< 12 ppm).} 
	
%\noindent
%We identify two possible improvements to the previous studies.
%Firstly, t
Excluding \citet{Chen2021-Persistenceofflaredriven}, the above studies are performed with one-dimensional photo- and thermochemical kinetics codes, thus ignoring longitudinal and latitudinal chemical exchange due to climate dynamics.
Close-orbiting, tidally locked exoplanets are expected to host superrotating equatorial jet streams that reach velocities of several kilometres per second \citep{Showman2002-Atmosphericcirculation, Showman2011-EquatorialSuperrotationonTidallyLockedExoplanets}.
Additional evidence for such a jet stream is inferred from many hot Jupiter phase curves where the brightest spot is shifted eastwards with respect to the substellar point \citep{Knutson2007-Mapofthedaynight, Komacek2017-AtmosphericCirculation, Zhang2018-PhaseCurvesofWASP33b}.
%
%with respect to the substellar point, that is observed in  (Knutson et al. 2007, Komacek et al. 2017, Zhang et al. 2018), is additional, observational evidence for such a jet stream.
%
%
%This has been inferred directly from phase curve observations of hot Jupiters, where the hottest point is shifted up to 50$\degrees$ east of the substellar point 
%
%
%\{[They play a role in mixing chemicals, and limb asymmetries]}
%Such jet streams have been shown to impact the distribution of heat \citep{Amundsen2016-TheUKMetOffice}, atmospheric constituents \citep{CooperShowman2006-DynamicsandDisequilibrium, Drummond2020-Implicationsofthreed} and aerosols \citep{Lines2018-SimulatingTheCloudy, Parmentier2013-3Dmixing} throughout the planet, which results in several spatial asymmetries that cannot be captured by one-dimensional codes.
When efficient enough, atmospheric dynamics can induce horizontal spatial variation in the distribution of chemical constituents \citep{CooperShowman2006-DynamicsandDisequilibrium,Mendonca2018-3DCirculationDrivingChemical, Drummond2020-Implicationsofthreed} and aerosols \citep{Parmentier2013-3Dmixing, Komacek2019-VerticalTracer, Steinrueck2019-TheEffectof3DTransport, Christie2021-Theimpactofmixingtreatments}, which cannot be captured by one-dimensional codes.
Subsequently, spatial asymmetries in the chemical and physical state of the atmosphere can give rise to differences between the morning and evening limb transmission spectra.
For example, the evening limb is typically hotter (and thus more vertically extended) and resembles more the dayside composition, owing to eastwards transport from the dayside, while the morning limb is colder and polluted with nightside constituents \citep{Baeyens2021-GridofPseudo2D}.
Taking into account these effects has been demonstrated to impact the final transmission spectrum \citep{Caldas2019-Effectsofafully3D}.
Although flares only affect the dayside, vigorous circulation can redistribute chemical species throughout the atmosphere.
Subsequently, this can affect the terminator regions which are probed with transmission spectroscopy.
Therefore, we need to consider a multi-dimensional treatment when studying the atmosphere's response to a flare event.
%
%\{[Flares impact dayside, we observe limbs]}
%Subsequently, superrotation is theorized to be responsible for a plethora of asymmetries between the morning and evening limb in transmission, although this is still observationally out of reach.
%Eastward heat transport yields a temperature contrast between both limbs, which causes the vertically extended evening limb to contribute more than opacity than the colder morning limb.
%Additionally, high meridional jets can pollute the cloudless dayside with condensates from the colder nightside through the morning limb and clear the evening terminator from any condensates.
%Finally, also the atmospheric composition differs between both limbs as a result from the temperature differences and advective horizontal transport of chemical compounds.
%I is likely that atmospheric circulation will also play a role the response of an atmosphere to one or more flare events.

%Secondly, the diversity of flare prescriptions used by previous authors is limited.
Stellar flares are known to span a wide range in energy, duration, emission line profile, occurrence frequency, peak flux, time evolution, etc. \citep{Schaefer2000-Superflares, Walkowicz2011-WhitelightFlares, Davenport2012-Multi-wavelengthCharaterization, Loyd2016-MUSCLESTreasurySruveyIII, Davenport2016-TheKeplerCatalogofStellarFlares, vanDoorsselaere2017-StellarFlaresObserved, Loyd2018-TheMUSCLESSurveyV, Loyd2018-HAZMETIV, Jackman2021-Stellarflaresfrom,  Jackman2021-Stellarflaresdetected,  Howard2021-NoSuchThingAsaSimpleFlare}.
The GFE of AD Leo, used by \citet{Segura2010-TheEffectofaStrong}, \citet{Venot2016-Influenceof} and \citet{Tilley2019-ModelingRepeatedMDwarf}, is considered a superflare, with a total radiative output of $\sim$$\num{e34}$ erg.
These flares are thought to be rare, while lower-energy flares tend to occur much more frequently.
As part of the MUSCLES Treasury Survey, \citet{Loyd2018-TheMUSCLESSurveyV} used data of several M dwarfs to create an idealised flare model
%with accompanying FFD
that has been used to study the effects of both rare high-energy and common low-energy flares on exoplanets \citep{Chen2021-Persistenceofflaredriven, Louca2022+Theimpactof}.

%\citep{Hawley1991-TheGreatFlare, Schaefer2000-Superflares, Walkowicz2011-WhitelightFlares, Davenport2012-Multi-wavelengthCharaterization, Shibayama2013-SuperflaresonSolar, Osten2016-AVeryBrightVeryHot, Loyd2016-MUSCLESTreasurySruveyIII, Davenport2016-TheKeplerCatalogofStellarFlares, Vida2017-FrequentFlaring, vanDoorsselaere2017-StellarFlaresObserved,  Loyd2018-HAZMETIV,	Jackman2018-Ground-baseddetection, Froning2019-Ahotultravioletflare, Tu2020-SuperflaresonSolar, Muheki2020-Highresolutionspectroscopy, Muheki2020-PropertiesofflaresandCMEs, Osten2010-TheMouseThat,  Seli2021-ActivityofTRAPPIST1analog, Namekata2021-Probabledetection, MacGregor2021-DiscoveryofanExtremely, MacGregor2021-UltrashortDurationFlareFromProximaCentauri, Jackman2021-Stellarflaresfrom,  Jackman2021-Stellarflaresdetected, Bourrier2021-TheHubblePanCETprogram}.

%\citep{Walkowicz2011-WhitelightFlares} Energy, frequency, duration
%\citep{Vida2017-FrequentFlaring}      Time evolution
%\citep{z     energy, frequency

%Although M dwarf activity is particularly relevant for habitability on rocky exoplanets, it poses a greater relevance towards a general better understanding of the star-planet connection.
%JWST will be able to measure the transmission spectra of rocky and especially gaseous exoplanets to the 50? ppm level, which requires a solid theoretical understanding of physical processes and impacting factors that alter the spectra to this amount of more.

In this paper, we aim to address how stellar flares of a M dwarf impact the neutral chemical composition of a close-orbiting, tidally locked gaseous planet ($\Teff=800$ K) with a superrotation-dominated circulation pattern.
We explore a variety of stellar flares that differ in energy, duration and occurrence frequency.
Additionally, we also investigate how the fluctuating composition alters the spectral signature of the planet in transmission.
In Sect. \ref{sec:modelling strategy}, we discuss the general modelling strategy, including the chemical kinetics model with initial and boundary conditions and radiative transfer.
Section \ref{sec: stellar flares} explains in more detail how the stellar flares are constructed.
%Our results that consider the impact of one particular type of flare on two different planet temperatures and three types of host stars are presented in Section \ref{sec: planet parameters}.
In Sect. \ref{sec: results}, we present our results for different stellar flares and planets.
We end with a discussion in Sect. \ref{sec: discussion}, before concluding in Sect. \ref{sec: conclusions}.

\section{Modelling tools} \label{sec:modelling strategy}
%As part of the effort to explore a variety of flare prescriptions, and perform multiple simulations, while still keeping computational costs acceptable, we choose not to adopt a fully 3D approach but a pseudo-2D?

\subsection{Pseudo-2D neutral disequilibrium chemistry}
\label{subsec: modelling-pseudo2D}

We use and slightly adapt the pseudo-2D chemical kinetics model of \citet{Agundez2014-Pseudo2Dchemicalmodel} that takes into account photo- and thermochemistry, vertical mixing (by means of eddy diffusion) and horizontal transport induced by equatorial superrotation.
In the model, the three-dimensional circulation pattern of the planet is approximated by a uniform zonal wind, described by a characteristic wind speed $\vwind$, that advects chemical species eastwards around the equator.
In practise, this is realised by a one-dimensional, rotating column that integrates the coupled continuity-transport equations for every species over a fixed longitude- and pressure-dependent thermal structure and pressure-dependent eddy diffusion profile.
After an integration time of
\begin{equation}
\tcol =\frac{\Pwind}{n_l } \, \mathrm{,}
\label{eq: tcol-ColumnIntegrationTime}
\end{equation}
with wind period
\begin{equation}
\centering
\Pwind =\frac{2\pi\rplanet}{\vwind} \, \mathrm{,}
\label{eq: Pwind}
\end{equation}
where $n_l$ is the number of longitudes and $\rplanet$ the planet radius, the column longitudinally shifts eastwards and again integrates the equations with updated initial conditions (advected chemical constituents, temperature structure and incident irradiation angle) specific for that longitude.
Because the longitudinal spatial dimension is effectively time-dependent, this formalism is dubbed pseudo-2D.
Although this model oversimplifies the complex three-dimensional dynamical structure of tidally locked exoplanets, it allows for relative computationally inexpensive use of chemical kinetics in a two-dimensional context.
After several column rotations, the longitude- and pressure-dependent chemical composition converges to a periodic steady-state in which the molar fractions are constant with respect to longitude.
We note that the wind period $\Pwind$ not only regulates the time it takes to complete one full column rotation but is in fact the same mathematical expression as $\tauzonal$, the zonal advection time-scale \citep[e.g.][]{Drummond2020-Implicationsofthreed}, with a constant value for the otherwise pressure-dependent zonal wind speed $u$.
However, $\tauzonal$ should be considered an order-of-magnitude estimate for the zonal advection while $\Pwind$ is an exact quantity associated with the pseudo-2D chemistry model.

Several studies have used a pseudo-2D formalism to investigate the two-dimensional, periodic steady-state molecular distributions of exoplanet atmospheres \citep{Agundez2014-Pseudo2Dchemicalmodel, Venot2020-GlobalChemistryandThermal, Baeyens2021-GridofPseudo2D, Baeyens2022-GridIIphotochemistry}.
However, in order to incorporate stellar flares into this framework, a time-dependent incident flux must be taken into account, and hence it is not guaranteed that a periodic steady-state will be reached.
%the incident stellar flux must become time-dependent and no periodic steady-state will be reached.
Instead, the chemistry model can be used to asses the dynamical response of the atmosphere to a perturbation of the incident radiation flux.
Therefore, we have adjusted the initial implementation of \citet{Agundez2014-Pseudo2Dchemicalmodel} such that the stellar flux can vary between column integrations.
However, we note that all boundary conditions, including the stellar flux, remain fixed for the duration of one column integration ($\mathrm{t}_{\mathrm{col}}$).

\begin{figure*}
	\centering
	\includegraphics[width=17cm]{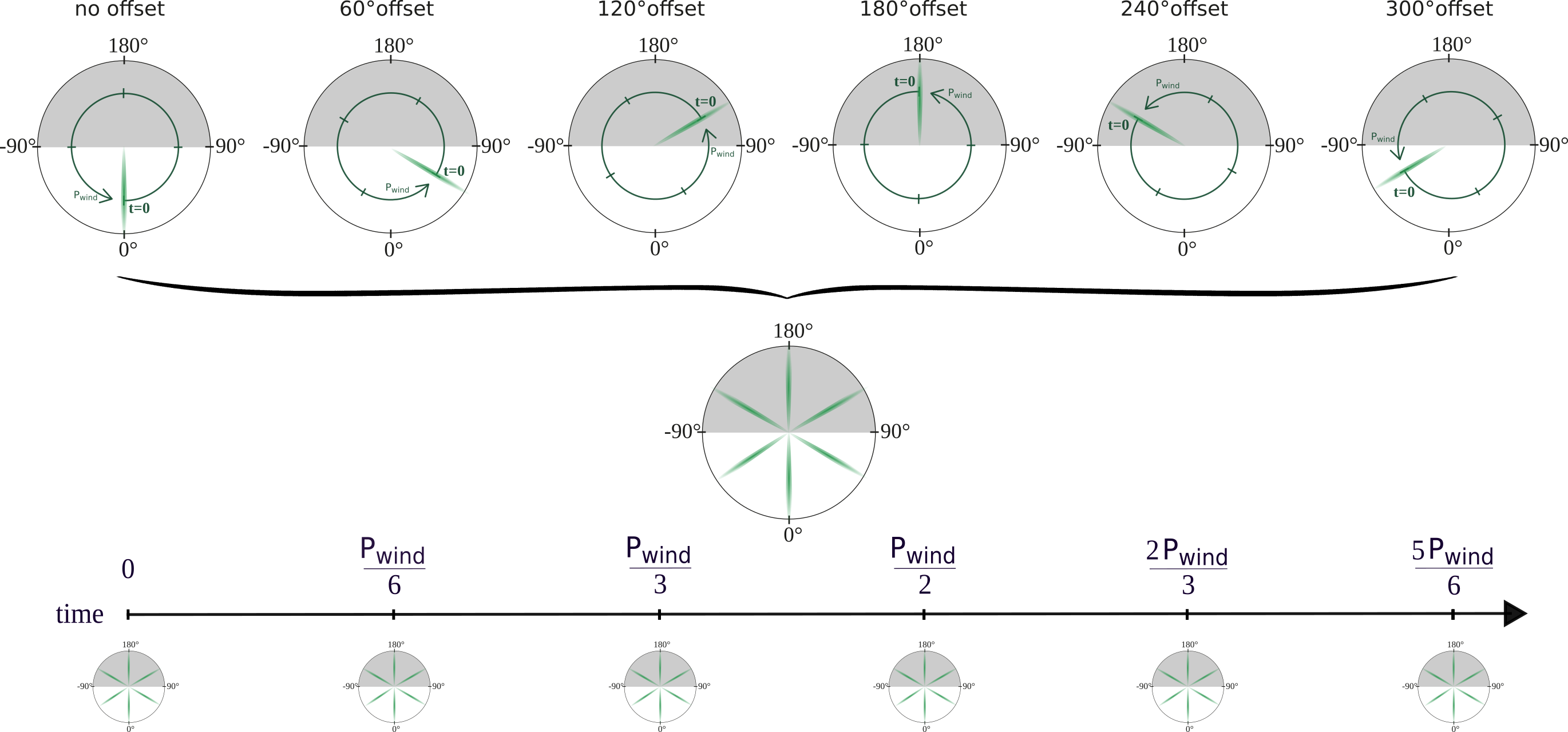}
	\caption{Schematic diagram that illustrates how individual pseudo-2D chemistry models are combined into a single two-dimensional distribution. First, six ($n^\prime_l=6$) pseudo-2D chemistry simulations with the same initial and boundary conditions are ran with an offset ($0\degrees$, $60\degrees$, ..., $300\degrees$) for the longitude which is first integrated ($t=0$). Subsequently, the information of all $n^\prime_l$ models are combined to a full two-dimensional distribution with time-sampling  $\Delta t =\Pwind/n^\prime_l$. For the models constructed in this paper, we adopt $n^\prime_l=90$.}
	\label{fig: S2-multimodelillustration}
\end{figure*}

By using the pseudo-2D chemistry code in a time-dependent manner, the time-longitude dependence puts a substantial constraint on the time sampling $\Delta t$.
Each longitude is only integrated every $\Pwind$ ($=\Delta t$), which typically amounts to several days for Jovian-sized planets (e.g. Eq. \ref{eq: Pwind} with $\vwind \sim$$\kms$).
Such poor time sampling would heavily restrict us in probing the atmosphere's response during and after a flare event.
Therefore, we combine several pseudo-2D models and construct a fully two-dimensional, time-dependent atmosphere with higher time sampling.
Our approach is illustrated in Fig. \ref{fig: S2-multimodelillustration}.
The longitude which is integrated first ($t=0$) in the pseudo-2D simulation was implemented by \citet{Agundez2014-Pseudo2Dchemicalmodel} to coincide with the substellar point but in fact can be any arbitrary longitude.
%In the classical use of the pseudo-2D code, such as implemented  by \citet{Agundez2014-Pseudo2Dchemicalmodel}, this starting point of the simulation ($t=0$) coincides with the substellar point.
By starting the simulation on a different longitude, the time-longitude dependence ensures that we obtain information of different longitudes at the same time-steps as before.
Therefore, we can run $n^\prime_l$ pseudo-2D models with the same initial and boundary conditions but apply an offset to the starting longitude.
By combining the information of all $n^\prime_l$ models on every time-step, we obtain a full two-dimensional time evolution of the atmosphere rather than the evolution of a single longitude as is the case for a single pseudo-2D model.
Furthermore, the time-sampling of the combined atmosphere model, containing $n^\prime_l$ individual models, now becomes $\Delta t =\Pwind/n^\prime_l$, which significantly improves the ability to study flare events on timescales shorter than several days.
%Essentially, once one assigns the starting point ($t=0$) to a certain longitude, the time-longitude dependence fixes the moments where the composition at each longitude is computed.
%
%
%By assigning the starting point to a longitude other that the substellar point, but keeping the other input parameters fixed, one performs the same simulation as before but now knows the composition at each longitude at different times.
%
%By applying these offsets on $n^\prime_l$ of the same pseudo-2D models and combining them into one two-dimensional distribution, the time sampling improves to $\Delta t =\Pwind/n^\prime_l$.
We note that the approach of combining separate models is justified because the very premise of the pseudo-2D formalism is that the chemical composition on longitude $l_i$ at time $t_i$ can be computed from information of only $l_{i-1}$ at $t_{i-1}$ and is thus independent from longitude $l_{i-1}$ at $t_i$.

\subsection{Chemical network}
\label{subsec: modelling-chemicalNetwork}

To compute the production and loss rates in every integration, we have used the chemical network of \citet{Venot2020-Newchemicalscheme} that involves 108 molecular species consisting of C, H, N, and O, linked by 1906 neutral-neutral reactions.
\citet{Venot2020-Newchemicalscheme} is an updated version of the network presented in \citet{Venot2012-Achemicalmodel}, which originated from the field of combustion chemistry and has been validated for temperatures and pressures relevant for hot Jupiters.
The network is accompanied by 52 temperature-independent photoabsorption cross-sections \citep{Venot2012-Achemicalmodel, Hebrard2013-PhotochemistryofHydrocarbons, Dobrijevic2014-CouplingofOxygenNitrogen}.
For further details regarding the chemical network such as lists of included molecules, reactions, and sources of the experimental data, we refer to \citet{Venot2012-Achemicalmodel} and \citet{Venot2020-Newchemicalscheme}.

\subsection{Pre-flare thermal and chemical state}
\label{subsec: modelling-preflarestate}

\begin{figure}
	\includegraphics[width=8.5cm]{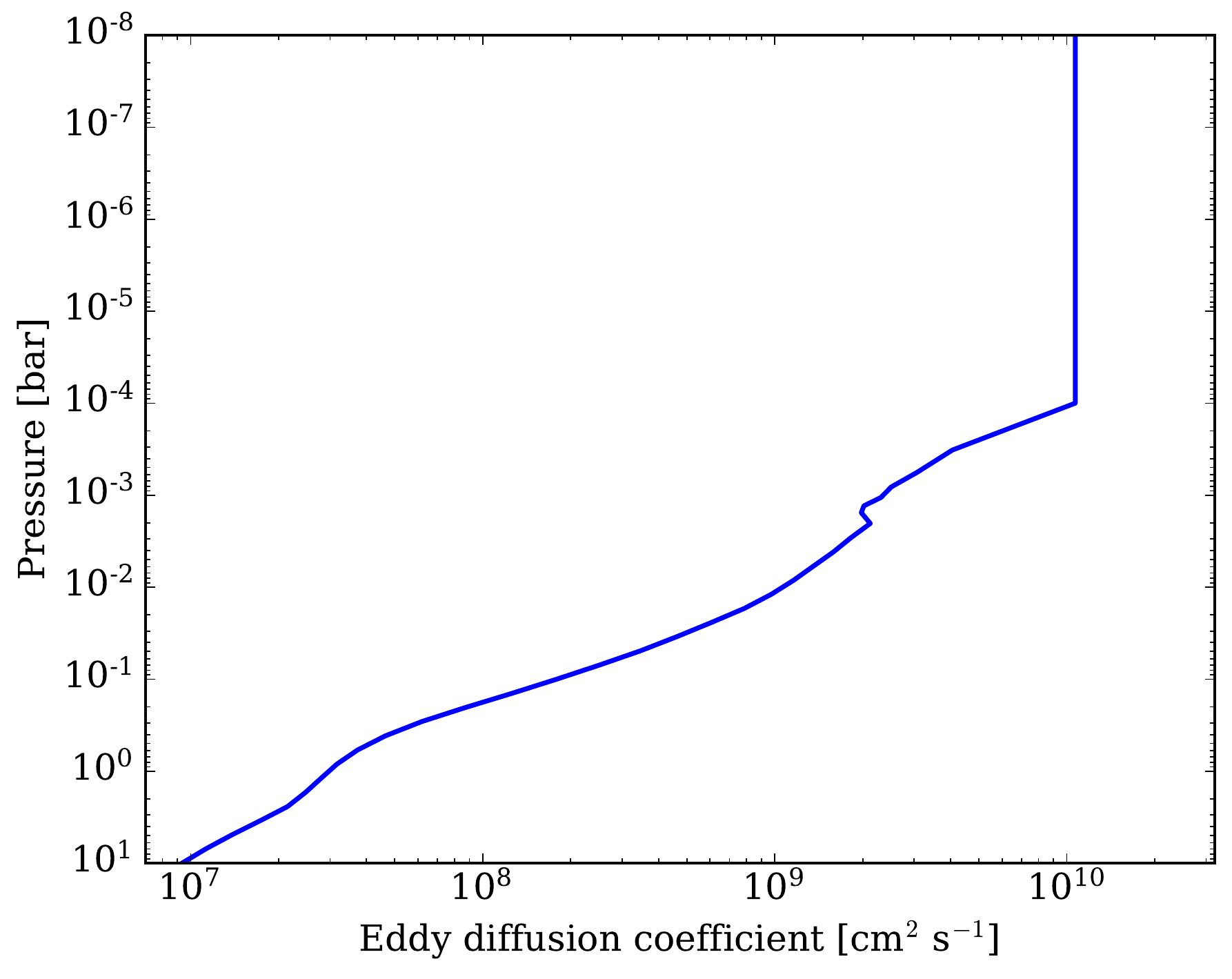}
	\includegraphics[width=8.5cm]{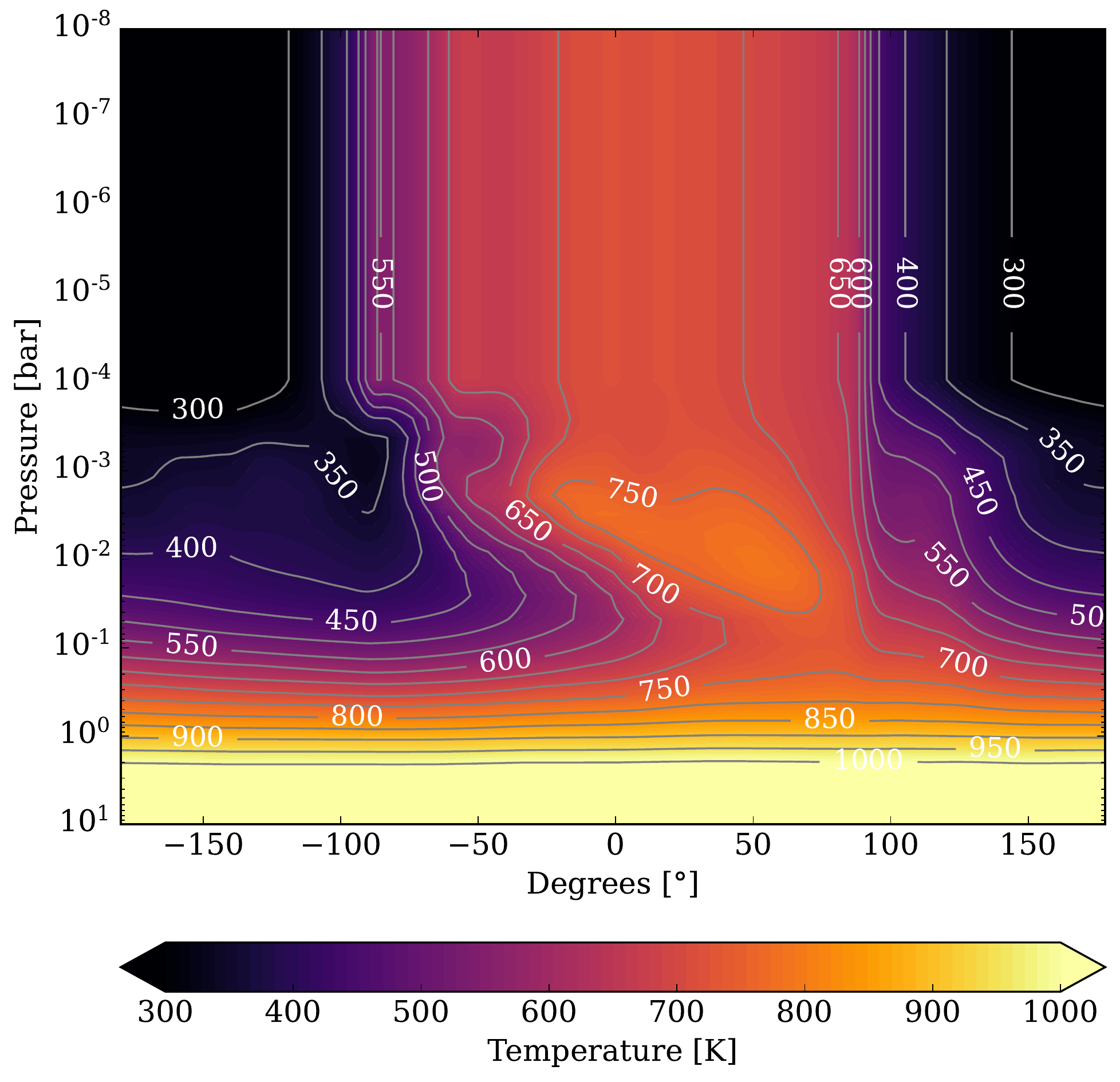}
	\caption{Temperature and vertical mixing information of the $\Teff =800$ K planet that orbits a M5-type star considered in this work. Data generated by \citet{Baeyens2021-GridofPseudo2D}. \textit{Upper:} Eddy diffusion coefficients $\Kzz$ with a constant value below $0.1 \mbar$. \textit{Lower:} Equatorial temperature structure with an isothermal upper atmosphere below $0.1\mbar$. $0\degrees$ denotes the substellar point. For pressure above $10$ bar, the temperature increases with pressure to values above $\num{1000}$ K so we have cut off the colourbar scale.}
	\label{fig: S2-thermalEddyWindofM800}
\end{figure}
\citet{Baeyens2022-GridIIphotochemistry} (B22 hereafter)
%\footnote{B22 is a follow-up study of \citet{Baeyens2021-GridofPseudo2D} that focusses on the impact of photochemistry on their grid of chemistry models. 
%At the time of writing, B22 is \textit{in review} but all necessary information regarding the climate and chemistry modelling can also be found in \citet{Baeyens2021-GridofPseudo2D}.}
have computed climate and pseudo-2D photochemistry models for a grid of tidally locked gaseous exoplanets that covers a range of effective temperatures ($\Teff \in \left[400, 600, ...,  2600 \mathrm{K} \right] $), surface gravities ($g \in \left[\num{e2}, \num{e3}, \num{e4}\right] \cmspersquared$), and spectral types of host stars (M5, K5, G5, F5).
From their study, we use the two-dimensional thermal profile, pressure-dependent eddy diffusion coefficients $\Kzz$, and zonal mean wind speed $\vwind$ of the model that corresponds to a  $\Teff=800$ K, $g =\num{e3} \cmspersquared$ planet orbiting a M5-type star.
The temperature profile (lower panel, Fig. \ref{fig: S2-thermalEddyWindofM800}) was extracted from a three-dimensional general circulation model output between 200 and $\num{e-4} $ bar.
B22 considered an equatorial band between latitudes $-20\degrees < \phi < 20\degrees$, from which they extracted a latitudinally weighted average that represents the equatorial thermal structure.
From $\num{e-4}$ bar down to $\num{e-8}$ bar, the atmosphere is assumed isothermal with respect to pressure and adopts the temperature value of the layer with $\num{e-4}$ bar.
We refer to B22 for a discussion on the limitations of this isothermal extension.

%The static temperature background itself (Figure \ref{fig: S2-thermalEddyWindofM800}), was assumed to be isothermal with respect to longitude below $0.1 \mbar$ \citep{Baeyens2021-GridofPseudo2D}.
%Also, the extracted equatorial temperature distribution \citep{Baeyens2022-GridIIphotochemistry} have already discussed that the temperature around 0.01 - 1 $\mubar$ regime can rise significantly due to strong irradiation absorption and ionization, while they keep all pressures below 0.1 $\mbar$ isothermal in their and this work.
%Although they test the effect of such high-altitude temperature rise, and conclude that it does not significantly affect the composition above 1 $\mubar$, it might impact our results?

%\textcolor{red}{Mention here that upper layers can a lot hotter, like addressed in B22?}
%\textcolor{green}{Should I say something about t-p profile itself? e.g. hotspot is at 10 mbar, and shifted with X degrees. Day-night contrast is xx kelvin... etc - the reader should normally just read B22 for interpretation of this}

%B22 computed the pressure-dependent eddy coefficients $\Kzz$  between 200 and $\num{e-4}$ bar by scaling the prescription by \cite{Komacek2019-VerticalTracer} with a factor 0.1, yielding
%\begin{equation}
%\Kzz(p) =0.1 \frac{w^2(p)}{\left(\frac{1}{\tauchem} + \frac{\mathrm{v}_\mathrm{hor}(p)}{\rplanet}\right)} \, \mathrm{,}
%\label{eq: Kzz(p)-Baeyens21}
%\end{equation}
%where $\mathrm{v}_\mathrm{hor}(p)$ is the isobarically averaged, horizontal wind speed ($\vwind$ here), $\tauchem$ the chemical timescale, and $w(p)$ the vertical wind speed.

We adopt the same pressure-dependent eddy diffusion coefficients ($\Kzz$) as in B22 (upper panel, Fig. \ref{fig: S2-thermalEddyWindofM800}).
The profile between 200 and $\num{e-4}$ bar is computed by using the prescription of \citet{Komacek2019-VerticalTracer} \citep[for more details, see][]{Baeyens2021-GridofPseudo2D}.
For the upper layers ($\num{e-4}$ to $\num{e-8}$ bar), B22 adopt a constant value of $\Kzz$ at $\num{e-4}$ bar.
The strength of the equatorial superrotating jet stream, that is the zonal mean wind speed $\vwind$, is computed to be $1.529 \, \kms$ from the zonally averaged wind speeds by averaging between $-20\degrees < \phi < 20\degrees$ and $10$ bar $> \mathrm{p} > 0.1 \mbar$.

For the pre-flare, periodic steady-state atmospheric composition, we repeat B22's pseudo-2D simulation with a vertical resolution of 60 and an increased longitudinal resolution ($n_l=180$).
We note that the elemental gas mixture is of solar metallicity and solar carbon-to-oxygen ratio (C/O = 0.55).
We fix the orbital, stellar, and planetary parameters to the values used by B22.
This encompasses the orbital separation of $0.01185$ $\AU$, planet mass $\mplanet =0.74$ $\mplanet$, stellar radius $\rstar =0.269$ $\rsun$, and planet radius $\rplanet =1.35$ $\rjup$.
Together with $\vwind$, the latter value leads to a wind period of $\Pwind \simeq \num{3.99e5} \, \mathrm{seconds} \simeq 4.6 \, \mathrm{days}$ (Eq. \ref{eq: Pwind}).
%In total, we run the simulation with above initial conditions for about one month  after the flare event.

In order to construct the fully two-dimensional atmosphere model, we run 90 ($n^\prime_l$) individual pseudo-2D models with each a longitudinal resolution of 180 ($n_l$) for about one month ($\sim\num{2.4e6}$ seconds) after the flare event and combine them following the method explained in Sect. \ref{subsec: modelling-pseudo2D} (see also Fig. \ref{fig: S2-multimodelillustration}).
%\{Note that $n^\prime_l$ does not equal $n_l$.}
This implies that the actual integration time of a column ($\tcol$) in an individual pseudo-2D model is twice as high as the time-sampling of the combined two-dimensional distribution ($\Delta t =\Pwind/n^\prime_l$).
We note that the step of combining individual models is done in post-processing and does not affect the simulations themselves.
%\{In other words, the numerical integration is performed with a higher time-sampling than the time-sampling with which the }
We choose $n^\prime_l=90$ to strike a balance between the computational expense of $n^\prime_l$ runs and a sufficiently high longitudinal resolution ($n_l$) of individual models.
Furthermore, $n^\prime_l$ has to have a common denominator with $n_l$ in order to successfully apply the combing strategy explained in Sect. \ref{subsec: modelling-pseudo2D} as otherwise, the various pseudo-2D models would probe different longitudes at the fixed time-steps and cannot be combined to a full two-dimensional atmosphere model.

\subsection{Synthetic transmission spectra}
\label{subsec: modelling-petitRADTRANS}

%petitRADTRANS
We computed synthetic transmission spectra during and after the flare event between 0.5 $\um$ and 30 $\um$ by using petitRADTRANS \citep{Molliere2019-petitRADTRANS} in low resolution mode ($\lambda/\Delta\lambda =1000$).
We include line absorption opacities of $\acetylene$, $\COtwo$, $\CO$, $\OH$, $\HCN$, $\NHthree$, $\methane$, $\water$, and $\Htwo$, Rayleigh scattering opacity of $\Htwo$ and $\He$, and continuum opacity from collision induced absorption by $\Htwo$-$\Htwo$ and $\Htwo$-$\He$.
Because atomic species other than hydrogen, carbon, nitrogen, and oxygen do not participate in the chemical disequilibrium modelling, we do not include species such as sodium and potassium in the radiative transfer calculations, despite their observed presence in gaseous exoplanet atmospheres \citep[e.g.][]{Redfield2008-SodiumAbsorptioninHD189733b, Charbonneau2002-DetectionOfExtraSolarAtmosphere, Nikolov2015-HSThotJupitertransmission,Chen2018-TheGTCexoplanettransitspectroscopysurveyIX}.
We also omit aerosol opacity sources in the radiative transfer computations to isolate effects of the flare on the chemical composition.

We use petitRADTRANS to acquire one-dimensional transmission spectra of the evening and morning limb separately.
This means that we extract a column from the longitude-temperature profile (Fig. \ref{fig: S2-thermalEddyWindofM800}) and chemical distribution at the relevant longitude, from which we compute the transit depths.
We adopt a reference pressure of 0.01 bar for a planet radius $\rplanet$ ($=1.35 \, \rjup$), from which the transit depths are computed.
We note that the adopted pseudo-2D formalism only represents the equatorial regions ($-20\degrees < \phi < 20\degrees$) while the entire terminator contributes to the transmission spectrum.
Our calculated spectra are therefore biased towards the chemical and physical state of the equator.
%\textcolor{red}{does a single evening transmission spectrum of 100 ppm mean that a (morning+evening)/2 is less than 100 ppm? In other words, do we overestimate the transit depth variations? }
%Recent studies have also shown the importance of a full three-dimensional treatment of the radiative transfer as the hotter, and thus vertically extended, dayside tends to screen the nightside, thereby dominating both limbs in transmission. \citep{Caldas2019-Effectsofafully3D}.
%
%
%
%Additionally, superrotation heats the evening limb and increases the scale height, which in turn affects the amplitude of the opacity features of the transmission spectra? (\textcolor{red}{Is it even a good idea to use Eq. \ref{eq: sec2-CombinedtransitDepth}?})
%\citep{MacDonals2020-WhyIsitSoColdinHere, Taylor2020-Understandingandmitigating}
%\\~\\
%While currently several 3D radiative transfer models are being developed (see Macdonald+21 for more info), we choose to make use of separate 1D models to probe differences in the limbs itself, without attempting to fit observations.
%\\~\\
%Why don't we include metals such as Na, K, ...?

\section{Stellar flares} \label{sec: stellar flares}

B22 have used the quiescent spectral energy distribution (SED) of GJ 876 from the MUSCLES Treasury Survey \citep{France2016-TheMUSCLESTreasurySurveyIMotivationandOverview, Youngblood2016-TheMUSCLESTreasurySurveyII, Loyd2016-MUSCLESTreasurySruveyIII, Youngblood2017-MUSCLESsurveyIV, Loyd2018-TheMUSCLESSurveyV} as an input for their atmospheric chemistry models with a M5-type host star.
GJ 876 is a M5 dwarf located at 4.7 pc \citep{France2016-TheMUSCLESTreasurySurveyIMotivationandOverview} with a temperature of $\Teff =3300$ K, surface gravity of $\log g =4.88 $ (cgs), and super-solar metallicity of [Fe/H] = 0.21 \citep{Rosenthal2021-TheCaliforniaLegacySurveyI}.
A total of four confirmed planets orbit GJ 876: two Jupiter-mass planets GJ 876b \citep{Marcy1998-AplanetaryCompanion} and GJ 876c \citep{Marcy2001-APairofResonant}, a long-period Neptune GJ 876e \citep{Rivera2005-APlanetOrbitingtheNearbyStarGJ876}, and a super-Earth GJ 876d \citep{Rivera2010-TheLickCarnegieExoplanetSurvey}.
The star itself was initially considered to be inactive, based on the absorption rather than emission of the $\mathrm{H}\alpha$ line.
%However, the Ca II H and K lines, which are more suited to quantify chromospheric activity \citep{GomesdaSilva2011-Longtermmagneticactivity}, 
However, additional observations have revealed high quiescent (X)UV emission \citep{Poppenhaeger2010-Coronalproperties, Walkowicz2008-CharacterizingtheNearUV} and frequent flaring \citep{France2012-TimeresolvedUltraviolet, France2016-TheMUSCLESTreasurySurveyIMotivationandOverview, Loyd2018-TheMUSCLESSurveyV}, which points to high stellar activity.
%\textcolor{red}{We use the same quiescent SED of GJ 876 as B22 and build upon it by constructing a simple, although realistic flare event that easily facilitates our aspiration to explore a variety of flares that differ in energy, duration, spectral coverage and occurrence frequency.}
We use the same quiescent SED of GJ 876 as B22 and construct a flare with a model developed by \citet{Loyd2018-TheMUSCLESSurveyV}.

%Observed flares: 				France+2012a
%X-ray and UV observations show active upper atmosphere: Walkowicz+2008, France+2012a
%High x-ray activity:			Poppenhaeger+2010, France+2016
%chromospherically inactive: 	Delfosse+1998
%high activity levels: France+2016
%Since low activity is typically associated with old stars ()
%
%-More info about this star, specifically its activity levels?- e.g. S index of 1.020+0.133 by Wright+2004

%\noindent

%Within the current state-of-the art chemistry models and the adopted model (time-resolution, uncertainties regarding reaction rates, photo absorption cross sections, mixing treatment, advection approximation for superrotation) in this work, it makes no sense to capture the complexity of a stellar flare event as a detailed time-series.
%Rather.

\subsection{Flare spectrum}
\label{subsec: StellarFlares-flareSpectrum}

In context of the MUSCLES survey, \citet{Loyd2018-TheMUSCLESSurveyV} have developed a Python package\footnote{\url{https://github.com/parkus/fiducial_flare}} to generate fiducial UV flares, with the intent of creating a consistent input for models that require time-dependent spectra \citep[e.g.][]{Chen2021-Persistenceofflaredriven, Louca2022+Theimpactof} and facilitate comparisons between them.
The fiducial flare spectrum is constructed with a spectral energy budget of both observed and reconstructed line emissions and a continuum source of a 9000 K blackbody \citep{Loyd2018-TheMUSCLESSurveyV}.
Furthermore, the time evolution profile of the flux is split up in a boxcar portion that encapsulates multiple, sustained peaks, followed by an exponential decay.
The spectral and temporal flare profile are normalised to the Si IV fluence such that one needs only the stellar quiescent Si IV flux to produce the flare in absolute units.

Given that $\tcol \simeq \num{2.2e3}$ seconds $\simeq\num{37}$ minutes (for $n_l =180$ and $\Pwind \simeq 4.6$ days, see Eq. \ref{eq: tcol-ColumnIntegrationTime}) is comparable and even longer than a typical flare duration, we are limited by the applied chemistry model and thus cannot capture a time profile that decays to quiescence during this time-span $\tcol$.
Therefore, we use the fiducial flare code to generate flares of various energies but make two major adjustments to the default values for the flare parameters.
First, we assume that all energy of the flare is contained in the boxcar portion of the model, thereby modifying the initial implementation where it is assumed that half of the flare energy is included in the exponential decay \citep{Loyd2018-TheMUSCLESSurveyV}.
Second, we adjust the scaling of the boxcar peak spectrum such that the time-span of the boxcar portion always equals $\tcol$ so that the time-profile does not evolve during a column integration $\tcol$.
We note that this time-span otherwise depends on the equivalent duration of the flare \citep[e.g. Fig. 20 in][]{Loyd2018-TheMUSCLESSurveyV}.
The resulting flares considered in this work thus always have a duration of $\tcol$ and consist only of a long peak, followed by an abrupt decrease to quiescence.
We explore the impact of a long exponential decay in Sect. \ref{subsec: Results-flare duration}.

%\{Time evolution}
%The time-evolution of the bulk of stellar flares can be described by an \textit{impulsive phase}, an initial rapid increase of the flux, followed by a \textit{gradual phase}, in which the flux returns to its quiescence levels \citep{Benz2010-PhysicalProcessesinMagnetically, Woods2011-NewSolarExtreme, Davenport2014-KeplerFlaresII, Benz2017-FlareObservations}.
%We are limited by the temporal resolution of the adapted chemistry model (Eq. \ref{eq: tcol-ColumnIntegrationTime}).
%Given that $\Delta t \sim \num{2.2e3}$ seconds $= \num{37}$ minutes for $n_l =180$ and $\Pwind = 4.6$ days, we sample the time-series with a time-resolution of $\Delta t$ as well.

\subsection{Total flare energy}
\label{subsec: StellarFlares-flareEnergy}

\begin{figure}
	%	\centering
	%	\includegraphics[width=17cm]{multi-model/FrankensteinIllustration.pdf}
	\resizebox{\hsize}{!}{\includegraphics{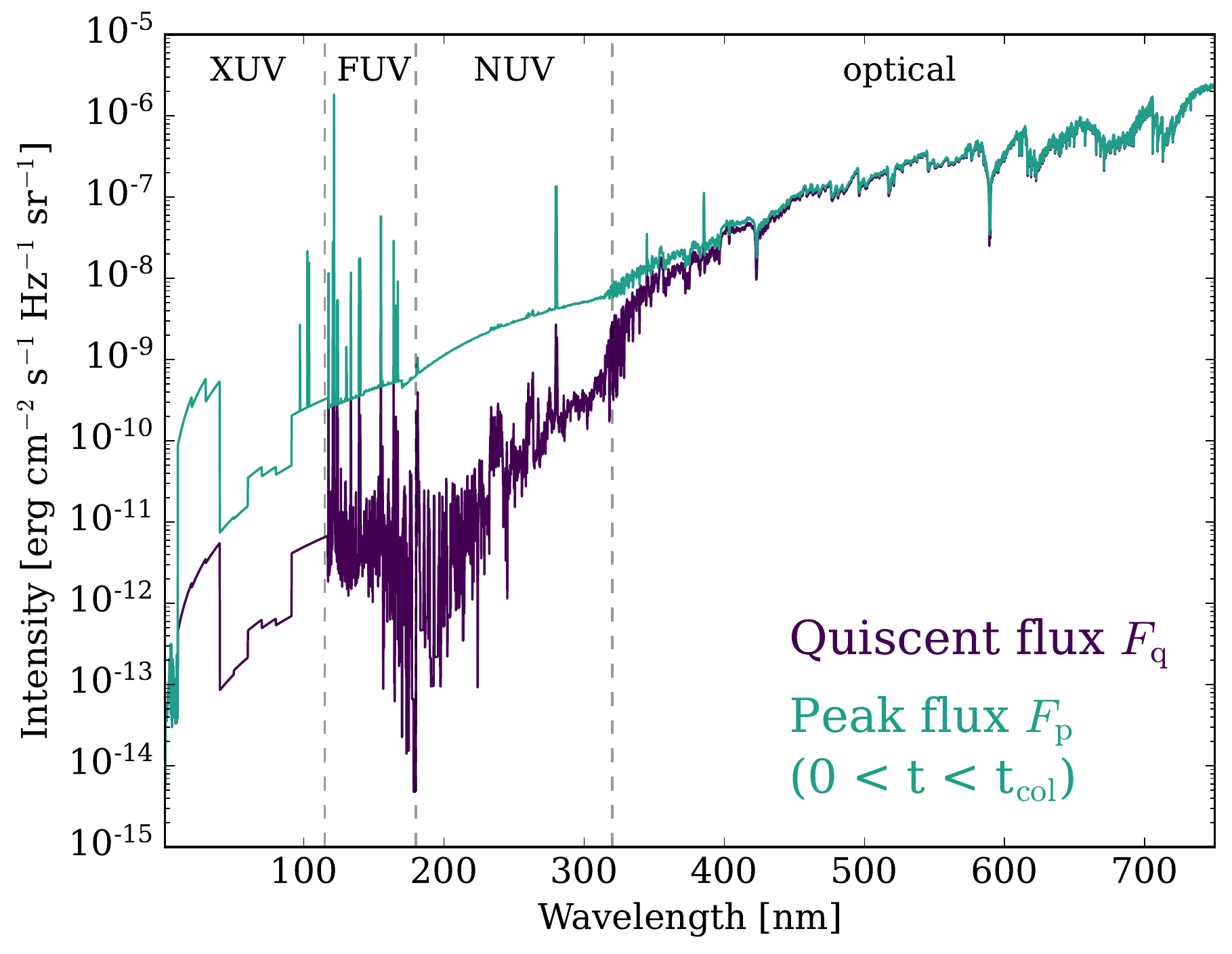}}
%	\caption{The constructed flare event of 4 hours and $E_\mathrm{tot}=\num{2e33}$ erg covering the FUV and NUV. Time $t=0$ marks the start of the flare event, where it reaches the peak flux $F_\mathrm{p}$ instantaneously. The time-series is sampled according to the time-resolution adopted in our individual pseudo-2D chemistry simulations ($\Delta t =\Pwind/180$).}
	\caption{Constructed flare event of $E_\mathrm{tot}=\num{2e33}$ erg, using the fiducial flare code of \citet{Loyd2018-TheMUSCLESSurveyV} with $\delta_\mathrm{Si \, IV} \simeq \num{e5}$ seconds. We assume an instantaneous increase to the peak flux $F_\mathrm{p}$ (green) for the duration of $\tcol \simeq \num{2.2e3}$ seconds, followed by an instantaneous decrease to the quiescent flux $F_\mathrm{q}$ (purple).}	
	\label{fig: stellarFlares-Flare}
\end{figure}

Flares cover a wide range of energies.
The total radiative energy of a stellar flare can be evaluated by integrating the flux difference between the flaring and quiescent state over time and wavelength:
\begin{equation}
E_\mathrm{tot} =4 \pi \rstar^2 \iint \left( F(\lambda,t) - F_\mathrm{q} (\lambda) \right) \, dt \,d\lambda \,\mathrm{.}
\label{eq: totalEnergyFlare}
\end{equation}
Additionally, the equivalent duration $\delta$ gives the time interval it takes for the quiescent star to emit the same amount of energy as was released during the flare ($E_\mathrm{tot}$) and yields
\begin{equation}
\delta =\frac{E_\mathrm{tot}}{4 \pi \rstar^2 F_\mathrm{q}} =\iint \left( \frac{F(\lambda,t) - F_\mathrm{q} (\lambda)}{F_\mathrm{q} (\lambda)} \right) \, dt \,d\lambda           \, \mathrm{.}
\label{eq: EquivalentDurationFlare}
\end{equation}
The GFE of AD Leo is estimated to have attained a total energy of $\sim$$\num{e34}$ erg \citep{Segura2010-TheEffectofaStrong, Loyd2018-HAZMETIV} across the entire measured wavelength domain.
In this work, we first consider a flare that is roughly equally energetic in equivalent duration for GJ 876, which results in $E_\mathrm{tot} =\num{2e33}$ erg.
Regarding radiative output in Si IV emission, the flare amounts to $\delta_\mathrm{Si \, IV} \simeq \num{e5}$ seconds.
We show both the quiescent and peak SED of the $E_\mathrm{tot} =\num{2e33}$ erg flare in Fig. \ref{fig: stellarFlares-Flare}.

A well-established relation exists between the flare energy and frequency of occurrence.
The flare frequency distribution (FFD) \citep[e.g.][]{Hawley2014-KeplerFlaresI, Davenport2016-TheKeplerCatalogofStellarFlares} shows that high-energy flares are much less common than low-energy counterparts. 
Such a FFD is described by a power law
\begin{equation}
d N(E) =k E^{-\alpha} d E \, \mathrm{,}
\label{eq: StellarFlares-FFDPowerlaw}
\end{equation}
where $N(E)$ is the number of flares in a given time interval, $E$ the flare energy, and $\alpha$ and $k$ constants.
Integrating this expression from $E$ to infinity gives
\begin{equation}
\log \nu  =C + \beta\log E =\log\left(\frac{k}{1-\alpha}\right) + (1-\alpha)\log E \, \mathrm{,}
\label{eq: StellarFlares-FFDLogExpression}
\end{equation}
where $\nu$ denotes the flare frequency for flares with energies of $E$ and higher.
\citet{Loyd2018-TheMUSCLESSurveyV} have found a striking similarity between the FFDs of inactive and active M dwarfs when computed with respect to equivalent duration instead of energy.
The same conclusion is reached when differentiating M dwarfs with age \citep{Loyd2018-HAZMETIV}.
When converted to equivalent duration, the hypothetical $\num{2e33}$ erg flare of GJ 876, considered in this work, would occur roughly every $10$ to $100$ days.
%, although this should be considered more an order of magnitude estimate
%\footnote{Note that \citet{Loyd2018-TheMUSCLESSurveyV, Loyd2018-HAZMETIV} compute equivalent durations in roughly 2 bands, 117-127 $\nm$ and 133 -143 $\nm$. We have accounted for this when computing the flare frequencies.}.
We note that a flare of $\num{2e33}$ erg, with an occurrence every $10$ to $100$ days, places this flare just in the superflare regime.
We further explore the effects of repeated flaring and in Sects. \ref{subsec: Results-repeated flaring} and \ref{subsec: Results-ffd}.

%Many other studies have constructed FFDs based on large surveys of both inactive and active M dwarfs, although with respect to total energy.
%Depending on the classification as inactive or active, and the observed wavelength region, these studies reveal a large scatter on the flare frequencies for a $\num{2e33}$ erg flare, ranging from less than 1 per hundreds of years to as high as 1 flare per day \citep{Davenport2012-Multi-wavelengthCharaterization, Hawley2014-KeplerFlaresI, Davenport2016-TheKeplerCatalogofStellarFlares,  Vida2017-FrequentFlaring,  Muheki2020-Highresolutionspectroscopy, Jackman2021-Stellarflaresdetected,  Seli2021-ActivityofTRAPPIST1analog}.
%\textcolor{red}{Loyd+18 and +20 have a lot of credibility in the literature. Should I therefore even mention the past state of the literature?}

\section{Results} \label{sec: results}

Throughout all the chemistry simulations performed in this study, we found that $\methane$, $\ammonia$, $\acetylene$, and $\HCN$ are the best indicators of flare-induced photochemistry.
This is in agreement with previous work that studied quiescent irradiation \citep[e.g.][]{Moses2011-Disequilibrium, Shulyak2020-Stellarimpact, Baeyens2022-GridIIphotochemistry}.
Additionally, these constituents contribute substantially to the transmission spectrum.
Therefore, we focus the discussion of our results on the evolution of the photochemically active species $\methane$, $\ammonia$, $\acetylene$, and $\HCN$ to characterise the effects of the flare on the atmosphere.
Other prominent constituents, such as $\water$, $\COtwo$, and $\CO$, are also affected but much less pronounced and are therefore only discussed briefly in Sect. \ref{subsec: Results-chemical composition} and appendices.
We express abundances of species $i$ in molar fraction, given by
\begin{equation}
\varphi (i) =\frac{n_i}{\sum_{j=1}^{N} n_j} \, \mathrm{,}
\end{equation}
where $n$ are number densities and $N$ is the total number of atmospheric constituents.

\subsection{Evolution of the chemical composition}
\label{subsec: Results-chemical composition}

\begin{figure*}
	\centering
	\includegraphics[width=17cm]{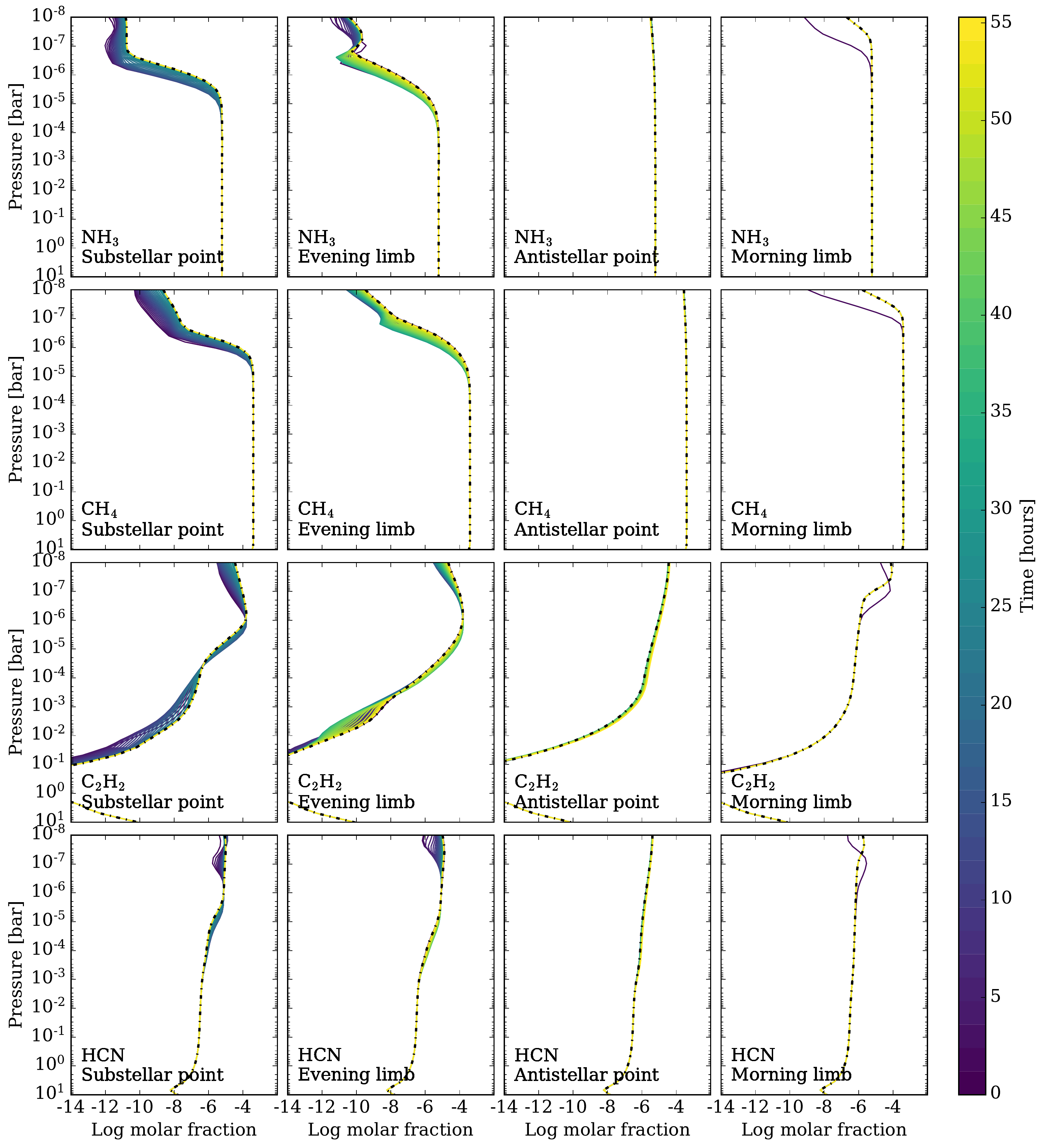}
	\caption{Evolution of $\methane$, $\ammonia$, $\acetylene$, and $\HCN$ (\textit{top to bottom}) on the substellar point, evening limb, antistellar point and morning limb (\textit{left to right}) during the first $\sim$$2.3$ days ($\Pwind/2$) after the flare event that started on $t=0$. The pre-flare distributions are represented by the black dashed-dotted lines.}
	\label{fig: results-chemComp-2.3daysevolutionOfLineProfiles}
\end{figure*}

Figure \ref{fig: results-chemComp-2.3daysevolutionOfLineProfiles} shows the evolution of $\methane$, $\ammonia$, $\acetylene$, and $\HCN$ at four longitudes (substellar and antistellar point, morning and evening limb) during the first $\sim$$2.3$ days ($\Pwind/2$) after the start of the flare event.
The latter time interval is the time it takes for horizontal transport to advect the constituents of the flared dayside completely to the nightside in our model (Eq. \ref{eq: Pwind}).
Naturally, the substellar point shows the largest variations in response to the flare event.
The upper atmosphere ($p < \num{0.1}\mbar$) becomes depleted in $\NHthree$ and $\methane$ by several orders of magnitude with respect to the pre-flare steady-state.
For example, at $1 \mubar$, $\varphi \left( \ammonia \right) / \varphi^{(\mathrm{ss})} \left( \ammonia \right) \simeq \num{6e-3}$ and $\varphi \left( \methane \right) / \varphi^{(\mathrm{ss})} \left( \methane \right) \simeq \num{4e-2}$ where $\varphi^{(\mathrm{ss})}$ denotes the pre-flare, steady-state molar fraction.
% NH3_pre-flare =1.3e-7 and NH3 =1.6e-11
% CH4_pre-flare =7e-05 and CH4 =2.8e-08
The dayside distribution of $\acetylene$ is affected throughout almost the entire atmosphere by depletion (p < 1 $\mubar$ and $0.1 \mbar$ < p < $\num{0.1}$ bar) down to a factor 0.1 at $\sim\num{0.1} \mubar$ and by a slight enhancement ($1 \mubar$ < p < $\num{0.1} \mbar$).
Also $\HCN$ is photodissociated in the upper layers (p < $1 \mubar$), reducing $\varphi (\HCN)$ by a factor $\sim$10 and slightly producing the constituent around $\sim$$10 \mubar$.

The evening limb distributions resemble the substellar point as a result of zonal advection that pollutes the terminator with the flared dayside.
However, the evening limb only returns to it's pre-flare composition after $\sim$$2.3$ days, while the substellar point recovers much faster.
Indeed, horizontal transport of un-flared nightside material reaches the substellar point after only $\sim$$28 \, \mathrm{hours}$ ($\Pwind/4$) after which the periodic steady-state composition is again recovered.
The evening limb remains in a perturbed state twice as long ($\Pwind/2$) as it is supplied with flared dayside constituents from as far back as the morning limb.
This also implies that the morning limb recovers very fast after the flare event.
Indeed, the morning limb returns to its pre-flare distribution in merely $\sim$2 hours after the start of the flare event.

Although the flare event alters the molar fractions of above species by orders of magnitudes on the substellar point and evening limb, the antistellar point remains fairly unaffected.
In a time interval of $\sim$$2.3$ days, the substellar point constituents have been advected to the antistellar point.
Upon entering the colder nightside, shielded from stellar irradiation, photodissociated methane and ammonia recombine and erase any traces of the initial depletion.
The same does not hold for the molar fractions of acetylene and hydrogen cyanide, two species that are known to be produced from photodissociated $\CHfour$ and $\NHthree$ \citep{Moses2014-Chemicalkinetics}.
The nightside becomes slightly enriched with $\acetylene$ and $\HCN$ up to a factor three, several days after the flare event.

\begin{figure*}
	\centering
	\includegraphics[width=17cm]{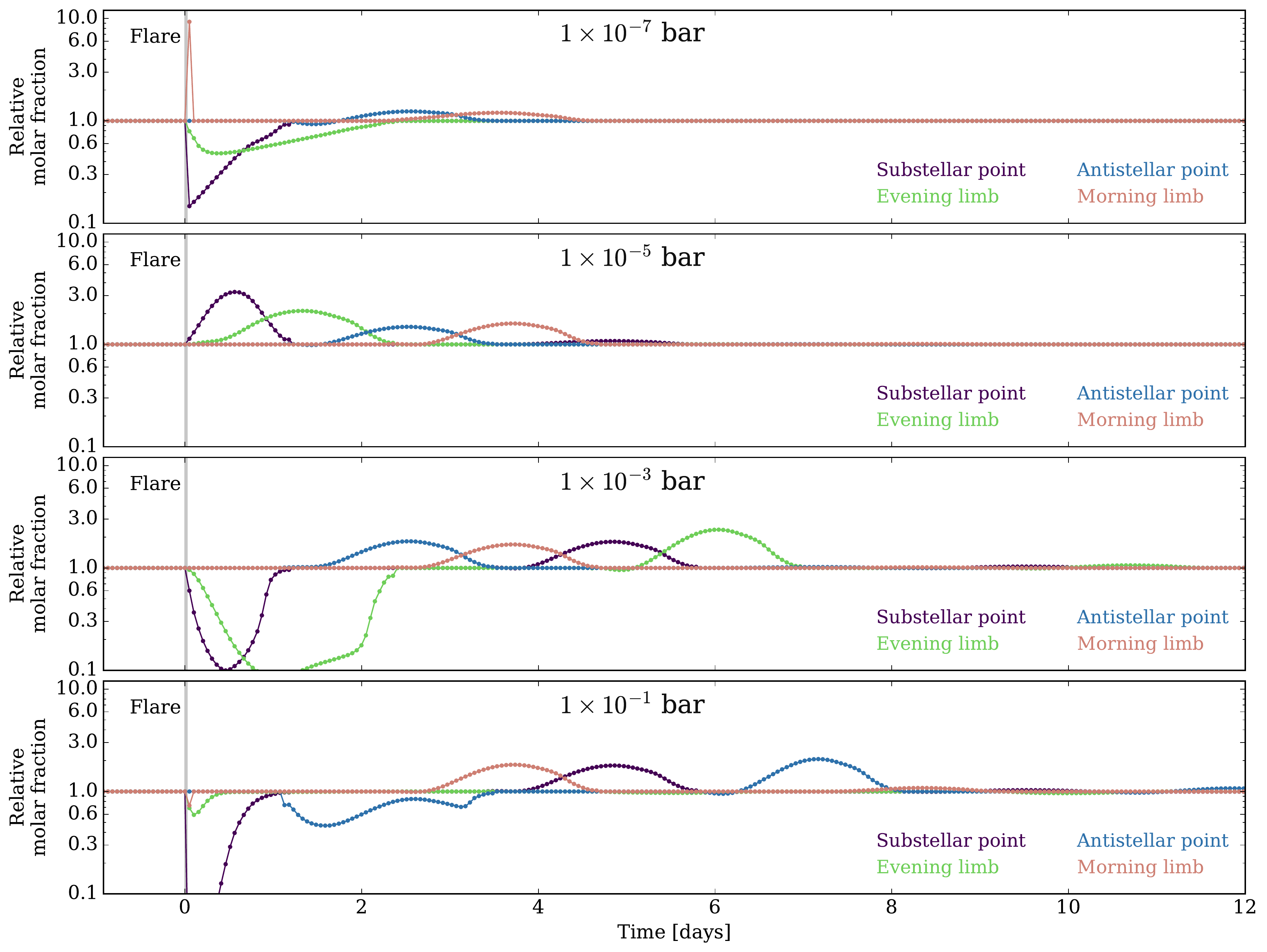}
	\caption{Evolution of $\acetylene$  in the first $\sim$2 weeks after the flare event at different pressures in the atmosphere on four longitudes. By plotting different longitudes, we trace the advection of atmospheric constituents by the horizontal wind in time.}
	\label{fig: results-chemComp-C2H2_evolution}
\end{figure*}

Although during the first $\sim$$2.3$ days after the start of the flare, the atmosphere's response is most extreme, subtle leftovers can be seen long after the flare event.
Figure \ref{fig: results-chemComp-C2H2_evolution} shows the molar fraction of $\acetylene$ relative to its pre-flare value at different pressures in the atmosphere up to two weeks after the flare event.
The nightside production of $\acetylene$ persists within the first $\Pwind$ time interval throughout the entire vertical atmosphere, with the exception of the upper layers ($0.1 \mubar$).
As the enriched nightside is advected further eastwards, the substellar point becomes enhanced by a factor 2 (after $\sim$4.6 days) and the evening limb by a factor 3 (after $\sim$5.8 days) in the middle regions ($10 \mubar$ and $1 \mbar$).
Furthermore, it is clear that the upper layers recover faster compared to regions of higher pressure.
Indeed, the deep atmospheric layer at 0.1 bar experiences variations on the nightside distribution for over a week after the flare event.

We briefly mention the effect of the flare event on species other than $\NHthree$, $\methane$, $\acetylene$, and $\HCN$ (see also Appendix \ref{app: Other species during the single flare event}).
In the upper layers ($p < 1 \mubar$) of the dayside, $\Htwo$, $\water$, and $\COtwo$ become slightly depleted and $\CO$ marginally enhanced by less than an order of magnitude.
The dayside hemisphere is enriched in atomic hydrogen by a factor ten up to pressures of $\sim$1 bar and slowly regains its pre-flare molar fractions within a factor two after  $\Pwind$.

\subsection{Evolution of the transmission spectra}
\label{subsec: Results-transmission spectra}
\begin{figure}
	\resizebox{\hsize}{!}{\includegraphics{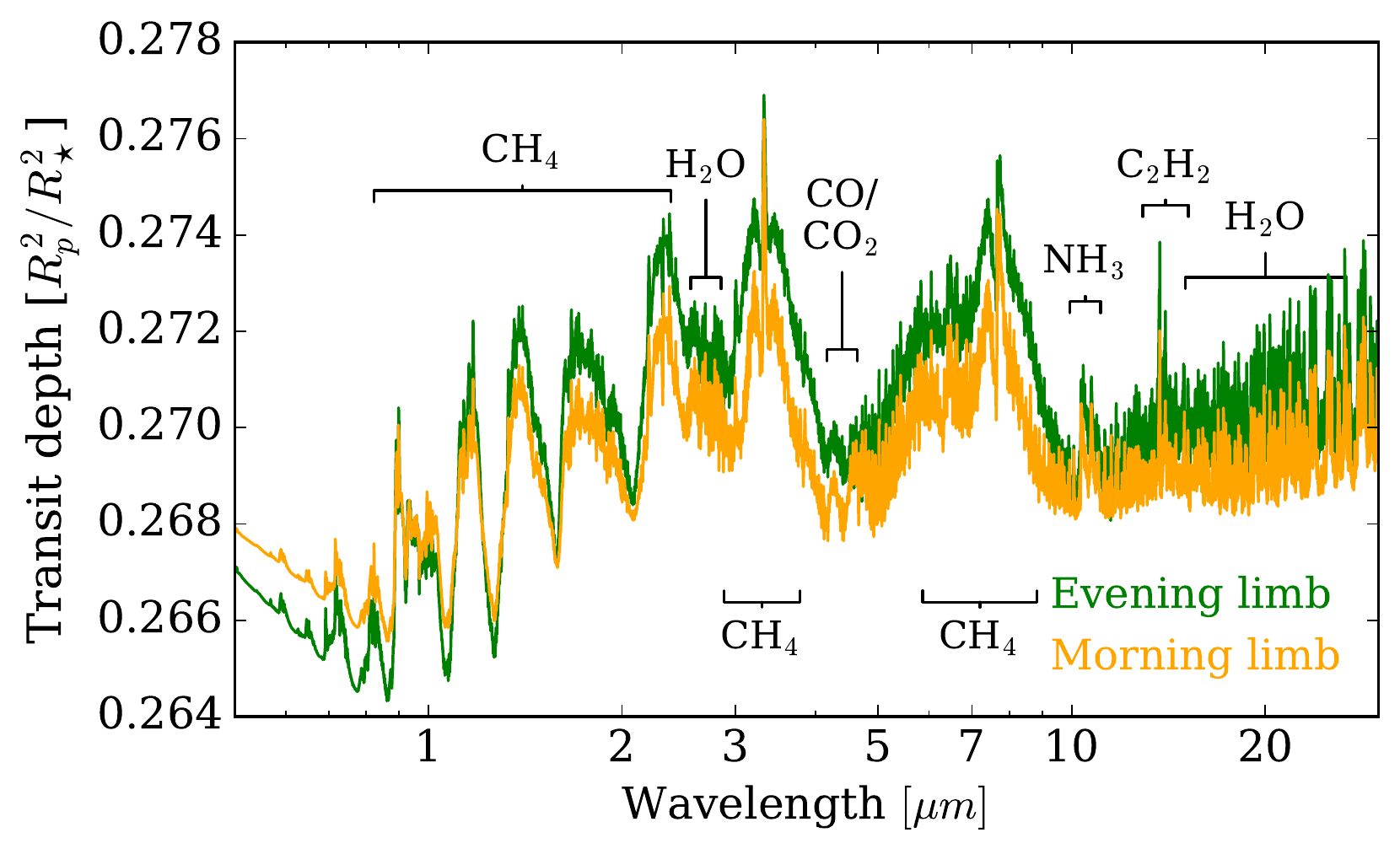}}
	\caption{Transmission spectra of the evening (green) and morning limb (orange) with the pre-flare molecular composition. The molecules that are predominantly responsible for certain absorption features are indicated on the relevant wavelength.}
	\label{fig: results-transSpec-preFlare}
\end{figure}

\begin{figure*}
	\centering
    \includegraphics[width=17cm]{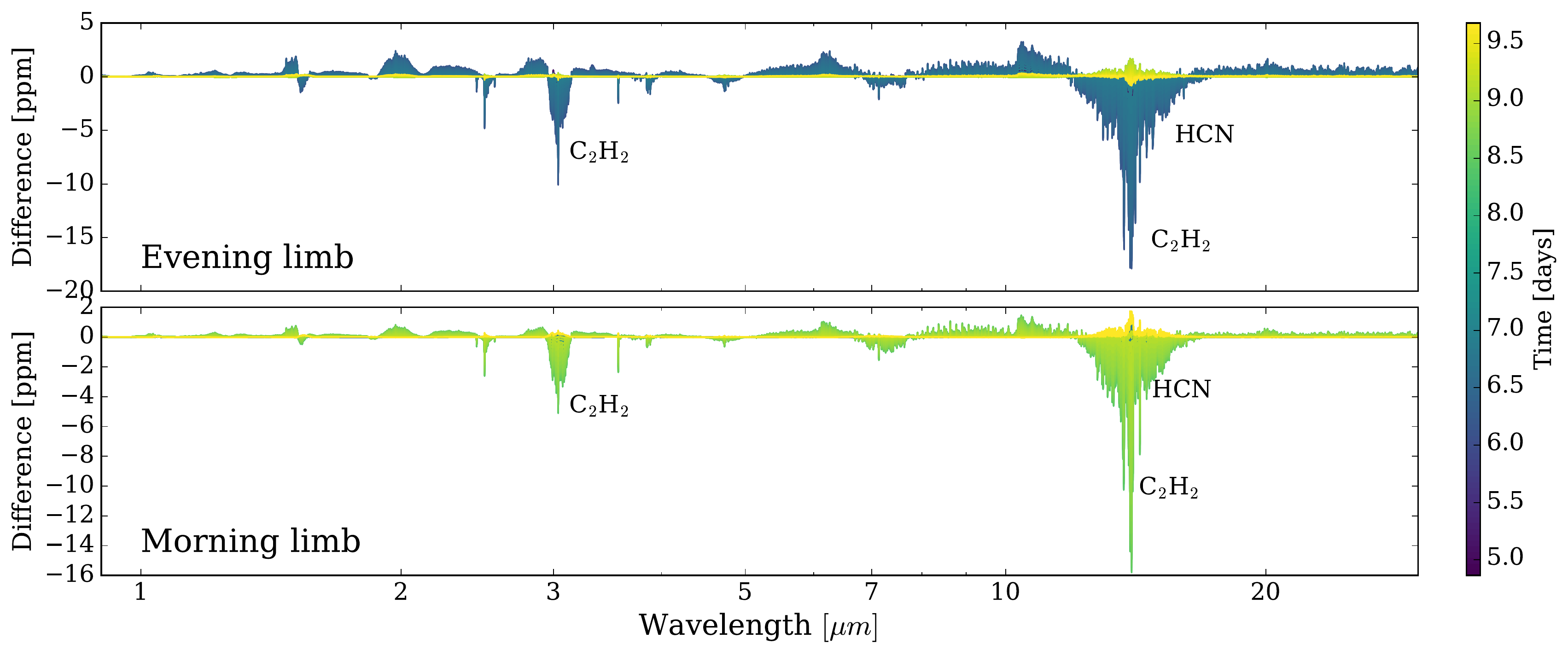}
	\includegraphics[width=17cm]{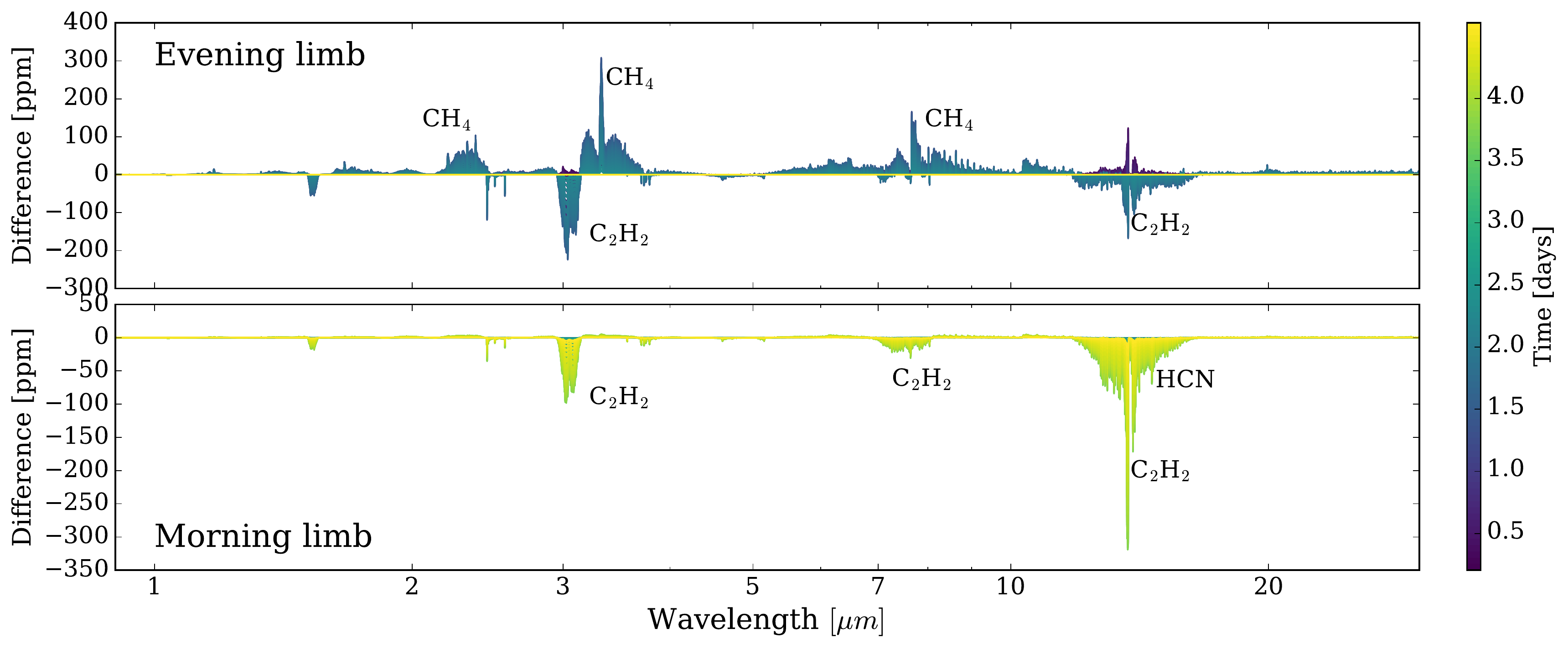}
	\includegraphics[width=17cm]{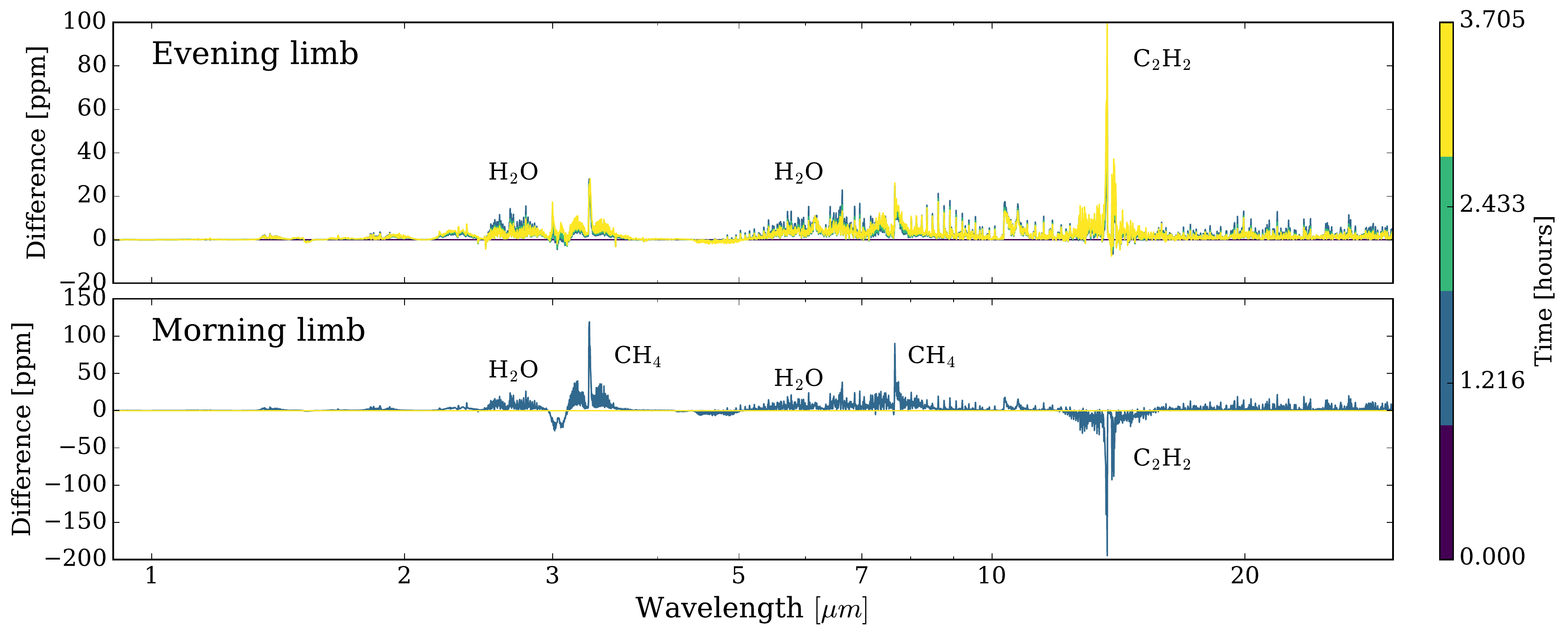}
	\caption{Differences in the transmission spectra of the evening and morning limb with respect to the pre-flare state during the flare (\textit{lower}), up to $\Pwind$ after the flare (\textit{middle}) and up to 2 $\Pwind$ after the flare (\textit{upper}).}
	\label{fig: results-transSpec-evolutionOfTrSpec}
\end{figure*}

In order to quantify the observational implications of variations in the chemical composition, we calculate transmission spectra during and after the flare event.
Figure \ref{fig: results-transSpec-preFlare} shows the transmission spectrum of the evening and morning limb for the atmosphere with a pre-flare molecular composition.
In Fig. \ref{fig: results-transSpec-evolutionOfTrSpec}, we compute spectra with atmospheric compositions perturbed by the flare (see Sect. \ref{subsec: Results-chemical composition}) and subtract these from the pre-flare spectrum.

During the first few hours after the flare ($0 < $ t $  \lesssim 37 $ min), the evening limb transit depths decrease by less than 30 ppm for $\lambda < 10 \um$, mainly due to $\water$ depletion (with small contributions from depleted $\CHfour$ and $\NHthree$).
At 14 $\um$, photodissociation of $\acetylene$ reduces the transit depth by $100$ ppm in a narrow spectral feature.
Additionally, depletion of $\HCN$ leads to a broader region around 14 $\um$ by less than 30 ppm reduction of the transit depth.
This is in contrast to the morning limb, where acetylene production creates additional opacity and raises the 14 $\um$ transit depth by 200 ppm.
Furthermore, methane depletion decreases spectral features around $3.3 \um$ and $7.8 \um$ by 50-100 ppm while $\water$ (and $\NHthree$) affect the transmission spectrum with less than 20 ppm. 
As expected from the effects of horizontal transport, the evening limb does not show signs of recovery $\sim$2 hours after the start of the flare, while the morning limb has returned back to its pre-flare state in this time interval.

During the first $\Pwind$ time interval, a large discrepancy exists between the evening and morning limb evolution in time.
As expected from the chemical response of the atmospheric constituents, the evening limb is perturbed up to $\sim$2.3 days after the flare event.
Methane depletion lowers the transit depths by up to 300 ppm (around $3.3 \um$ and $7.8 \um$) while $\acetylene$ production introduces additional opacity around 3 and 14 $\um$, altering the spectrum up to 200 ppm.
We note that the latter molecule now increases the transit depth on the evening limb, as opposed to the first few hours.
The morning limb remains free from deviations during the first $\sim$$2.3$ days, with the exception of the first hours after the flare.
Afterwards, from $\sim$$2.3$ to $\sim$$4.6$ days, the roles reverse and the evening limb experiences no substantial deviations from the pre-flare state because, in this time interval, the constituents of the nightside are advected completely to the dayside again, crossing the morning limb.
As shown in Sect. \ref{subsec: Results-chemical composition}, $\acetylene$ is produced substantially on the nightside.
Subsequently, this enhancement is responsible for increasing the transit depths at 3 $\um$ ($\sim$100 ppm) and 14 $\um$ ($<$ 350 ppm).
The latter absorption feature consists of a broad band ($\sim$100 ppm) due to HCN production accompanied by the strong narrow peak of $\acetylene$.
Finally, for t > $4.6$ days, $\acetylene$ (and $\HCN$) continues to temporarily enhance the transit depths, although by less than 20 ppm, when the flared dayside is again transported over the evening and morning limbs.

\subsection{Impact of different climates}
\label{subsec: Results-different climates}
Before exploring flares of different energies, we turn our attention to the factors that impact the climate of tidally locked gaseous exoplanets.
In particular, the strength of equatorial superrotation and temperature structure play a crucial role in determining the spatial, steady-state distribution of molecular species \citep{Baeyens2021-GridofPseudo2D}.
Therefore, we run two additional models with the same flare event of \num{2e33} erg. 
First, we double $\vwind$ to $3.058 \kms$, while keeping other parameters fixed (\textsc{run 1}).
Second, we consider another planet model of \citet{Baeyens2021-GridofPseudo2D} with $\Teff=1600$ K and set $\vwind=1.529 \kms$ (\textsc{run 2}), the latter equalling the value adopted in previous runs (see Sects. \ref{subsec: Results-chemical composition} and \ref{subsec: Results-transmission spectra}).
The model of \textsc{run 2} has a two-dimensional temperature structure with a much stronger day-night gradient and also adopts different, generally higher, eddy diffusion coefficients.
Additionally, the main carbon-bearing molecule in the pre-flare chemical composition is $\CO$ for models with $\Teff =1600$ K, compared to $\CHfour$ for $\Teff =800$ K.
An effective temperature of $1600$ K yields an orbital separation of $0.00296 \, \AU$, which affects the incident flux levels of stellar radiation.
In this section, we refer to the nominal $\Teff =800$ K model with $\vwind =1.529 \kms$, described in Sects. \ref{subsec: Results-chemical composition} and $\ref{subsec: Results-transmission spectra}$, as \textsc{run 0}.
Additional differences between the set-ups of these runs are summarised in Table \ref{tab: Results-Climates-Runs}.
For a detailed discussion on the differences between the pre-flare compositions of the models with $\Teff =800$ K (\textsc{run 0} and \textsc{run 1}) and $\Teff =1600$ K (\textsc{run 2}), we refer to B22.

\begin{table*}[t]
	\begin{center}
		\centering
		\caption{\label{tab: Results-Climates-Runs} Parameters for the models with different climate conditions. All runs were performed with the same stellar flare of $E_\mathrm{tot} =\num{2e33}$ erg. \textsc{run 0} was described in detail in Sects. $\ref{subsec: Results-chemical composition}$ and $\ref{subsec: Results-transmission spectra}$.}
		\begin{tabular}{lcccccc}
			\hline\hline\rule[0mm]{0mm}{5mm}
			Run &  $\Teff$ [K]  &  $a$ [AU]  & $\vwind$ [$\kms$] & $\Pwind$ [days] & $n^\prime_l$  & $\Delta t$ [s] \\
			\hline\hline\rule[0mm]{0mm}{5mm}\textsc{run 0}  & 800 & $\num{0.01185}$ & 1.529 & $\sim$4.6 & 90 & $\sim\num{2.2e3}$ \\
			\textsc{run 1}  & 800 & $\num{0.01185}$ & 3.058 & $\sim$2.3 & 45 & $\sim\num{2.2e3}$ \\
			\textsc{run 2}  & 1600 & $\num{0.00296}$ & 1.529 & $\sim$4.6 & 90 & $\sim\num{2.2e3}$ \\
			\hline\rule[0mm]{0mm}{3mm}
			%\multicolumn{2}{c}{$^a$ \footnotesize{This represents the stellar radius as it is set in the pseudo-2D simulation. }}
		\end{tabular}
	\end{center}
\end{table*}

\begin{figure*}
	\centering
	\includegraphics[width=17cm]{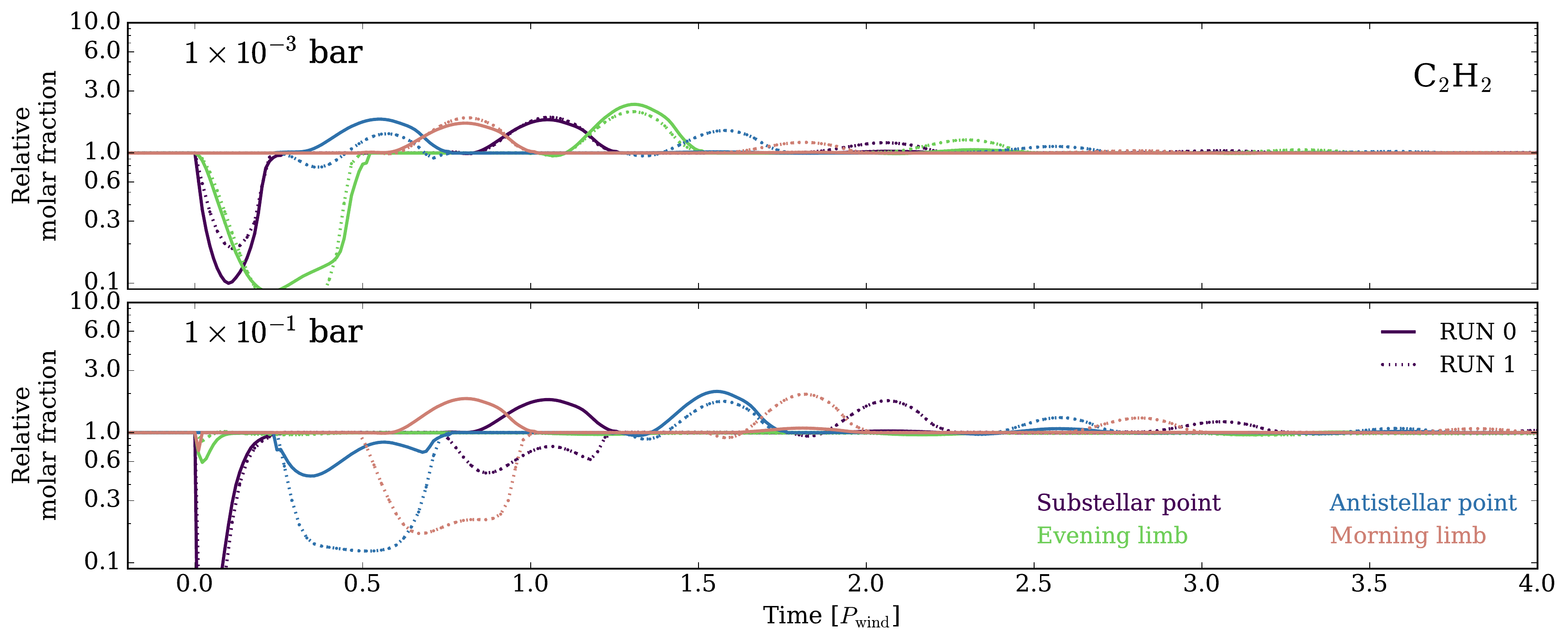}
	\caption{Evolution of acetylene on different longitudes in the middle ($1 \mbar$) and deep atmosphere (0.1 bar) for \textsc{run 0} and \textsc{run 1}. We note that the horizontal axis is in units of $\Pwind$, which differs between runs (see Table \ref{tab: Results-Climates-Runs}).}
	\label{fig: results-climates_run01}
\end{figure*}
\begin{figure*}
	\centering
	\includegraphics[width=17cm]{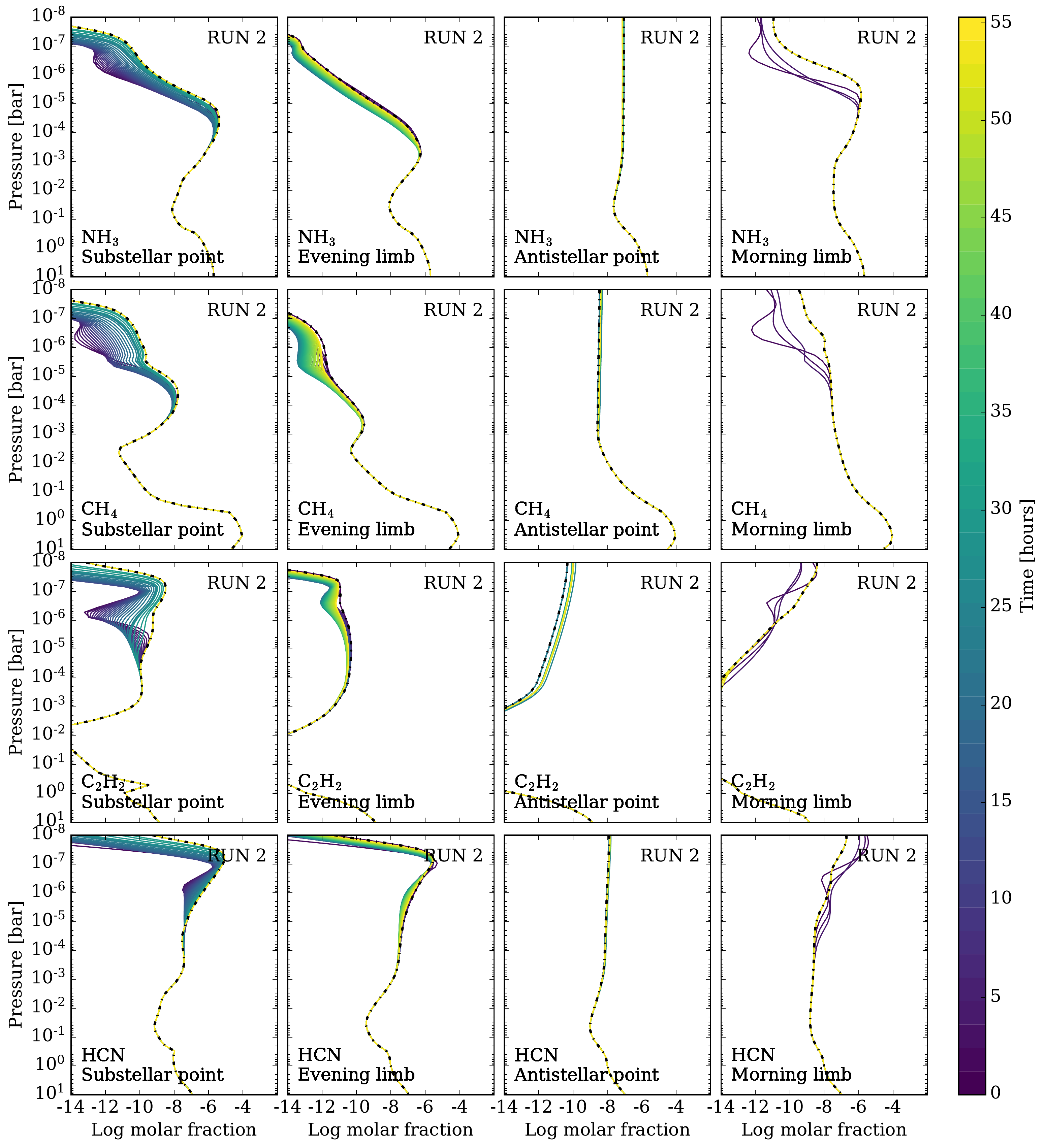}
	\caption{Same as Fig. \ref{fig: results-chemComp-2.3daysevolutionOfLineProfiles}, but for \textsc{run 2} (see Table \ref{tab: Results-Climates-Runs}).}
	\label{fig: results-climates_run2}
\end{figure*}

A selection of the results are shown in Figs. \ref{fig: results-climates_run01} and \ref{fig: results-climates_run2}.
%The response of the $\Teff =800$ K planet atmosphere with $\vwind =3.058 \kms$ is qualitatively very similar to the results presented in Section \ref{subsec: Results-chemical composition}.
%Due to a higher wind speed, the evolution in time is sped up with a factor two.
In \textsc{run 1}, again two phases can be identified, before and after $\Pwind/2$.
There are only minor differences related to the amount of methane and ammonia depletion on the dayside (t < $\Pwind/2$) and $\acetylene$ and $\HCN$ production on the nightside (t > $\Pwind/2$) (only $\acetylene$ is shown).
After multiple $\Pwind$ (which differs between \textsc{run 0} and \textsc{run 1}, see Table \ref{tab: Results-Climates-Runs}) and thus multiple advection periods around the planet, the molar fractions of $\acetylene$ also remain high compared to the pre-flare values.
However, because $\Pwind^{(\textsc{run 0})} \simeq 2 \, \Pwind^{(\textsc{run 1})}$, this does not imply that the atmosphere in \textsc{run 1} remained perturbed longer by the flare event.
Instead, because horizontal transport is twice as fast, and the vertical mixing and chemical timescales remains fixed, more column rotations can be completed in the same time interval before the pre-flare distribution is recovered (Fig. \ref{fig: results-climates_run01}).
Therefore, we conclude that, in the case of a stronger zonal wind, the constituents of the flared dayside are circulated more times around the planet but the atmosphere's response in time stays qualitatively the same.

For \textsc{run 2} (Fig. \ref{fig: results-climates_run2}), the atmosphere's response to the flare event is predominately controlled by the differences with \textsc{run 0} in orbital separation and planet temperature.
Because this hypothetical planet is about four times closer than in \textsc{run 0}, the photochemical rates during the flare event are higher.
Therefore, the depletion in the upper atmosphere of the dayside is much higher. 
More specifically; around $\sim$$1 \mubar$, $\varphi \left( \ammonia \right) / \varphi^{(\mathrm{ss})} \left( \ammonia \right) \simeq \num{e-3}$, $\varphi \left( \methane \right) / \varphi^{(\mathrm{ss})} \left( \methane \right) \simeq \num{3e-4}$, $\varphi \left( \acetylene \right) / \varphi^{(\mathrm{ss})} \left( \acetylene \right) \simeq \num{2e-4}$, and $\varphi \left( \HCN \right) / \varphi^{(\mathrm{ss})} \left( \HCN \right) \simeq \num{9e-2}$.
Furthermore, a higher effective temperature of 1600 K implies that chemical equilibrium determines the composition down to a few $\mbar$, while for $\Teff=800$ K (\textsc{run 0}) this was only the case for the deep layers (p > 1 bar).
Given the short chemical timescales, the flare event does not substantially affect the middle/deep atmosphere in \textsc{run 2}.
Aside from the differences originating in temperature and orbital separation, the atmosphere's response to the flare event is qualitatively similar as in \textsc{run 0} for the first $\sim$2.3 days ($\Pwind/2$) after the start of the flare.
Although we do not show the evolution of the constituents after $\sim$2.3 days, we report that the nightside becomes slightly enhanced in $\acetylene$, as was also the case in \textsc{run 0}.
However, we identify a striking difference in the evolution of the morning limb of \textsc{run 2}.
When the constituents of the flared dayside are advected over the nightside towards the morning limb, we see little to no enhancement of $\acetylene$ and $\HCN$ on the morning limb several days after the flare event.
This is in contrast to the evolution of the model atmospheres with $\Teff =800$ K, where we find a substantial production of these species on the nightside that are subsequently advected several times across the planet.
It thus seems that the hotter planet recovers its pre-flare composition in less than the $\Pwind$ time interval, and thus differs from the atmosphere's response of the planet with $\Teff =800 \, \mathrm{K}$.
Although many aspects can contribute to this difference, we accentuate that the chemical timescales are shorter in warmer regimes, which supports a faster recovery after a perturbation such as a flare event.
We discuss the evolution of the transmission spectra of \textsc{run 2} in Appendix \ref{app: Run 2 transmission spectra}.

\subsection{Impact of total flare energy}
\label{subsec: Results-flare energy}

\begin{figure*}
	\centering
	\includegraphics[width=17cm]{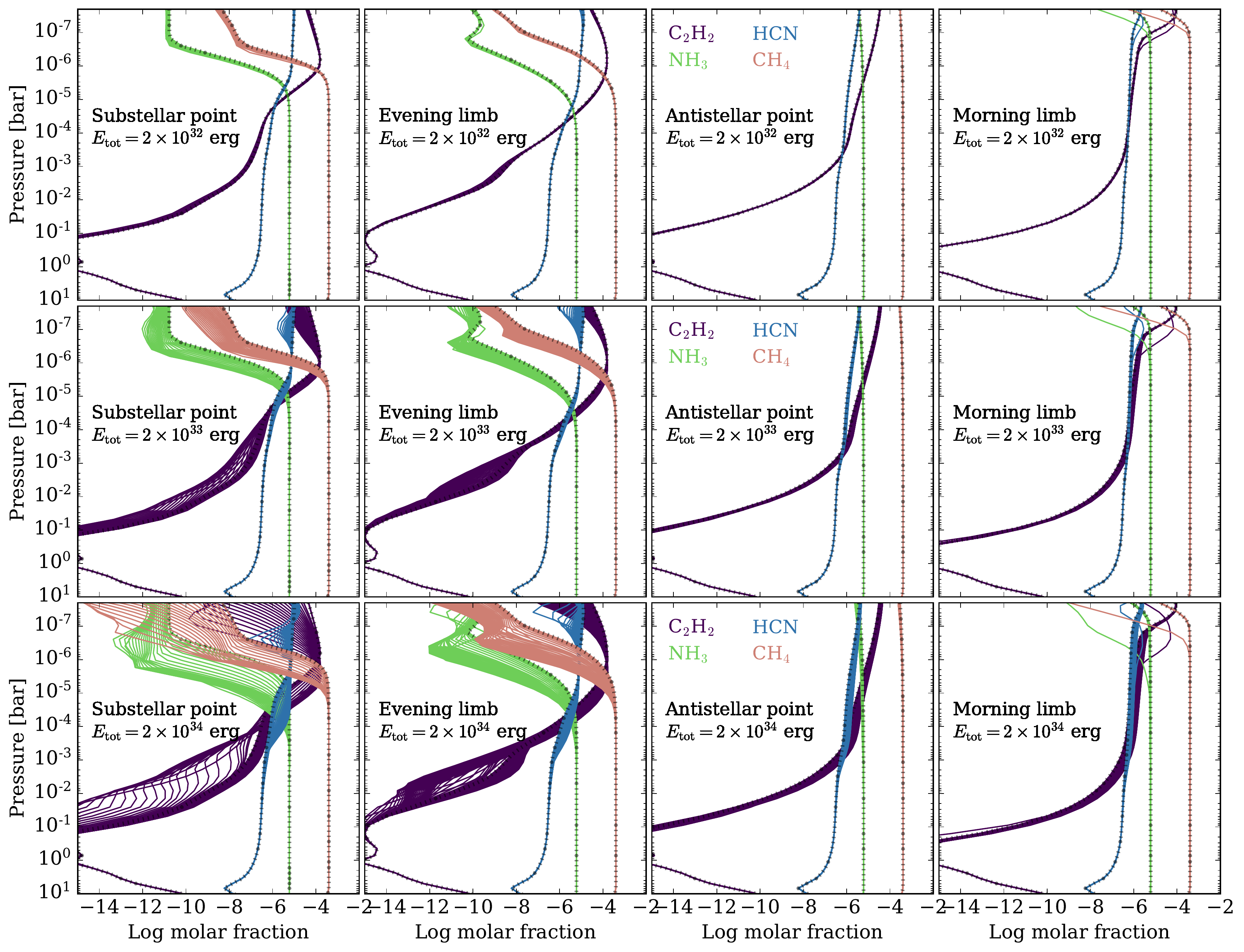}
	\caption{Distributions of $\CHfour$, $\NHthree$, $\acetylene$, and $\HCN$ during and $\sim$1 month after a flare event of total energy $\num{2e32}$ erg (\textit{upper row}), $\num{2e33}$ erg (\textit{middle row}) and $\num{2e34}$ erg (\textit{lower row}). \textit{From left to right:} substellar point, evening limb, antistellar point and morning limb. The pre-flare distributions are represented by the black dotted lines.}
	\label{fig: results-Energy}
\end{figure*}

We analyse the $\Teff =800$ K planet with $\vwind =1.529 \, \kms$, and run the chemistry model with flares of different energies.
%, while keeping other parameters such as flare duration and wavelength coverage fixed.
We consider flares of energies $E_\mathrm{tot} \in [\num{2e32}, \num{2e33}, \num{2e34}]$ erg, by scaling the peak flux $F_\mathrm{p}$ accordingly.
%For all three flares, we adopt $\eta =1.727$ so that $f(t =4 \, \mathrm{hours}) =\num{e-3} $ still holds (Eq. \ref{eq: time-evolutionChadney2017}) although this implies that $F(t =4 \,\mathrm{h}) / F_\mathrm{q}$ slightly differs among flares.
The simulation with $E_\mathrm{tot} =\num{2e33}$ erg will be used as a baseline to compare other simulations with different $E_\mathrm{tot}$ to, and we refer to Sects. \ref{subsec: Results-chemical composition} and \ref{subsec: Results-transmission spectra} for a detailed discussion on this baseline model.
Figure \ref{fig: results-Energy} shows the molar fractions of $\CHfour$, $\NHthree$, $\acetylene$, and $\HCN$ throughout the $\sim$1 month simulation domain.
As expected, the overall impact of the $\num{2e32}$ erg flare is small, while the opposite holds for the $\num{2e34}$ erg flare.
The latter causes more $\CHfour$ and $\NHthree$ depletion in the upper regions of the dayside and evening limb, as well as more production of $\HCN$ and $\acetylene$ on the nightside and morning limb.
For example, relative to the pre-flare composition, this quantifies as $\varphi \left( \ammonia \right) / \varphi^{(\mathrm{ss})} \left( \ammonia \right) \simeq \num{e-5}$ and $\varphi \left( \methane \right) / \varphi^{(\mathrm{ss})} \left( \methane \right) \simeq  \num{3e-5}$ on the substellar point at $1 \mubar$.
During the simulation with $E_\mathrm{tot}=\num{2e32}$ erg, the molar fractions of none of the species change by more than a factor 20 relative to their pre-flare values.
Overall, variation of $E_\mathrm{tot}$ results in a scaling of the depletion and enhancement patterns throughout the atmosphere compared to the case of $E_\mathrm{tot} =\num{2e33}$ erg.
Thus the atmosphere's response is qualitatively the same for flares of different energies.
%In other words, no additional chemical pathways substantially produce or deplete species with higher/lower flare energy.
Also the spectral features that are affected in the evolution of the transmission spectra are qualitatively similar with respect to the  case.
The maximal variation in transit depth on the evening limb during $0 < t \lesssim2.3$ days are 60 ppm and 800 ppm for flare energies of $\num{2e32}$ erg and $\num{2e34}$ erg respectively, compared to a value of 300 ppm for the baseline model of $E_\mathrm{tot}  = \num{2e33}$ erg.
For the morning limb during $2.3 \, \mathrm{days} \lesssim t \lesssim4.6$ days, these numbers are 60 ppm and 700 ppm for flare energies of $\num{2e32}$ erg and $\num{2e34}$ erg, compared to 350 ppm for the baseline model. 
%To reproduce the spectra of the three models considered, one could multiply the strength of de-/increase in transit depth (Figure \ref{fig: results-transSpec-evolutionOfTrSpec}) with factors 1/4, 1 and 2 for flare energies of $\num{2e32}$ erg, $\num{2e33}$ erg and $\num{2e34}$ erg.

\subsection{Impact of flare duration}
\label{subsec: Results-flare duration}

%\textcolor{green}{idea is here to only compare block flare to a flare that decays very long}

%\textcolor{red}{Don't forget to make the duration so that Flux(4h)=0.001 Fq and not f(4h)=0.001}

In Sect. \ref{subsec: StellarFlares-flareSpectrum}, we adapted the fiducial flare code of \citet{Loyd2018-TheMUSCLESSurveyV} such that all flares considered in this work last $\sim$37 min ($\tcol$) while in reality, the duration is correlated to the energy and can range from minutes to hours \citep[e.g.][]{Walkowicz2011-WhitelightFlares, Hawley2014-KeplerFlaresI,  Seli2021-ActivityofTRAPPIST1analog, Bourrier2021-TheHubblePanCETprogram, Jackman2021-Stellarflaresdetected}.
Therefore, we aim to explore if the flare duration has a substantial impact on the atmosphere's response by comparing the previous results to long duration flares.
We follow \citet{Chadney2017-Effectofstellarflares} in taking a simplified approach for the time evolution by considering an instantaneous increase to the peak flux $F_\mathrm{p}$, followed by an exponential decay towards quiescence $F_\mathrm{q}$.
The time-dependent stellar flux can then be expressed as
\begin{equation}
F(\lambda,t) =f(t)\cdot F_\mathrm{p}(\lambda) + (1 - f(t)) \cdot F_\mathrm{q}(\lambda) \, \mathrm{,}
\label{eq: time-evolutionChadney2017}
\end{equation}
with $f(t) =\exp(-\eta t)$ and $\eta$ a constant determined by the duration of the flare, assuming that the flare happens at $t=0$.
We note that the difference with the previous flare is that after the first column integration ($0 < t < \tcol$), the flux decreases exponentially to quiescence rather that instantaneously, although the exponential decay still happens in time-steps of $\tcol$.
%Although the bulk of the flares can be approximated by this simplistic parametrization, more complex patterns have been observed \citep[see e.g.][]{Hawley2014-KeplerFlaresI, Davenport2014-KeplerFlaresII, Vida2017-FrequentFlaring, Howard2021-NoSuchThingAsaSimpleFlare}.
%We do not consider such alternative time profiles in this study.
%The flare duration is technically defined as the time interval between deviation from and decay to quiescent flux levels.
%Because the final phase of the flare is hard to distinguish from the quiescent flux in observations, a full-width-half-maximum-like timescale $t_\mathrm{1/2}$ is used that measures the time after which the flux has dropped to half of its peak value \citep{Davenport2014-KeplerFlaresII, Jackman2021-Stellarflaresdetected}.
%
%
%
Following the duration of the GFE of AD Leo, we consider a duration of $\sim$4 hours by setting $\eta =1.727$ so that $f(t =4 \, \mathrm{hours}) =\num{e-3} $ (Eq. \ref{eq: time-evolutionChadney2017}).
To conserve an integrated total energy of $\num{2e33}$ erg for a duration of 4 hours, we scale the peak flux to $\sim$$65\%$ of its initial value (shown in Fig. \ref{fig: stellarFlares-Flare}).
This implies that the remaining $\sim$$35\%$ is included in the exponential decay.
Furthermore, we construct flares with a duration of 2 and 8 hours for which we adopt a value for $\eta$ of 3.454 and 0.863 respectively.
This implies that the peak flux must be $\sim$$83\%$ and $\sim$$40\%$ of the initial value (Fig. \ref{fig: stellarFlares-Flare}) for the flares of 2 and 8 hours respectively to amount to an energy of $\num{2e33}$ erg.

\begin{figure}
	\resizebox{\hsize}{!}{\includegraphics{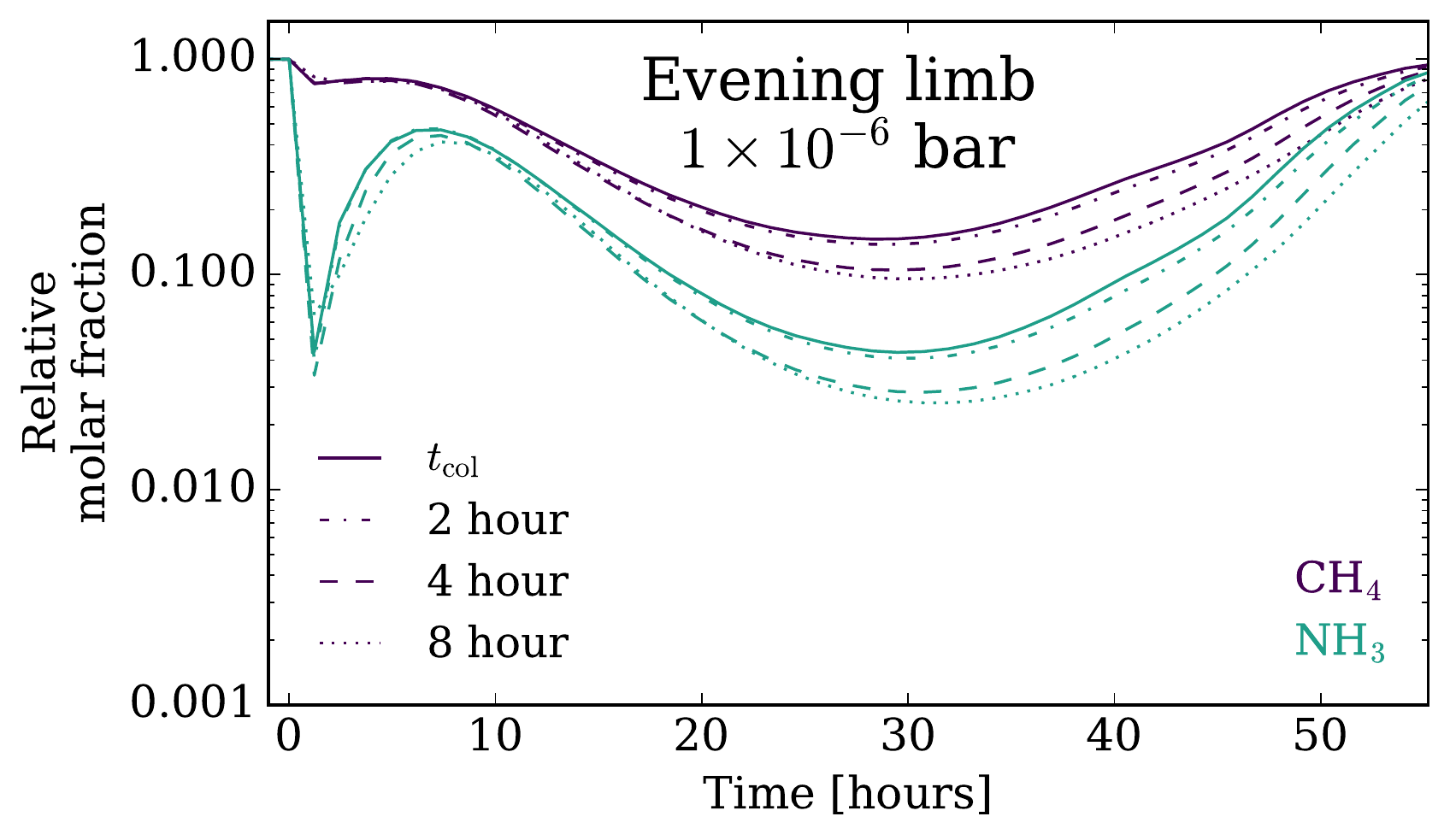}}
	\caption{Evolution of $\methane$ and $\NHthree$ during the first $\sim$2.3 days ($\Pwind/2$) after the start of the flare event in the upper atmosphere ($ 1\mubar$) of the evening limb. Three simulations are shown that correspond to $\num{2e33}$ erg flares of 2 (dashed-dotted), 4 (dashed) and 8 (dotted) hours. The simulation discussed in Sects. \ref{subsec: Results-chemical composition} and \ref{subsec: Results-transmission spectra}, with a flare of $\tcol$ without exponential decay, is shown by the solid line.}
	\label{fig: results-Duration-NH3CH4}
\end{figure}

%To obtain an integrated total energy of for all durations (2, 4 and 8 hours), we scale the peak flux $F_\mathrm{p}$ with factors 2, 1, and $\frac{1}{2}$ respectively and adopt 
%
%\textcolor{red}{83 \% of the OG peak for the 2-hour flare and 40 \% of peak for the 8-hour}
%
%A total duration of 4 hours amounts to $t_\mathrm{1/2} =24$ minutes, which is in agreement with a typical empirical energy-duration relation \citep[e.g.][]{Jackman2021-Stellarflaresdetected}.
%
%
%\{We fix the total radiative output of the flare to $E_\mathrm{tot} =\num{2e33}$ erg}, but now vary the duration i.e. the time it takes to reach $f(t) =\num{e-3} $.
%
%The differences between these runs are rather subtle.
%Therefore, we exemplify said differences by showing the evolution of $\methane$ and $\NHthree$ during the first $\tauzonal$ (4.6 days) after the flare event in the upper atmosphere ($ 1\mubar$) of the evening limb (Figure \ref{fig: results-Duration-NH3CH4}).

In Fig. \ref{fig: results-Duration-NH3CH4}, we show the evolution of $\methane$ and $\NHthree$ during the first $\sim$2.3 days ($\Pwind/2$) after the start of the flare event in the upper atmosphere ($ 1\mubar$) of the evening limb.
Although the differences between these runs are rather subtle, we find that the effect of the flare duration manifests itself in a number of ways.
Firstly, the curves corresponding to long duration flares (e.g. 8 hours) are elongated to the right on the horizontal time-axis, which results from a slower decrease to quiescence compared to shorter flares.
Secondly, flares with longer duration seem to deplete the evening limb more in $\CHfour$ and $\NHthree$ than shorter flares after advection across the dayside from $\sim$1 day onwards, although this effect seems limited to a factor two.
Interestingly, the lower value of the peak flux for longer flares does not seem to play a significant role compared to the duration of the exponential decay.
Due to the lack of diffusive mixing in the longitudinal direction in our models, the above effects do not dissipate and last throughout the entire simulation but remain relatively small.

\subsection{Impact of repeated flaring}
\label{subsec: Results-repeated flaring}

\begin{figure*}
	\centering
	\includegraphics[width=17cm]{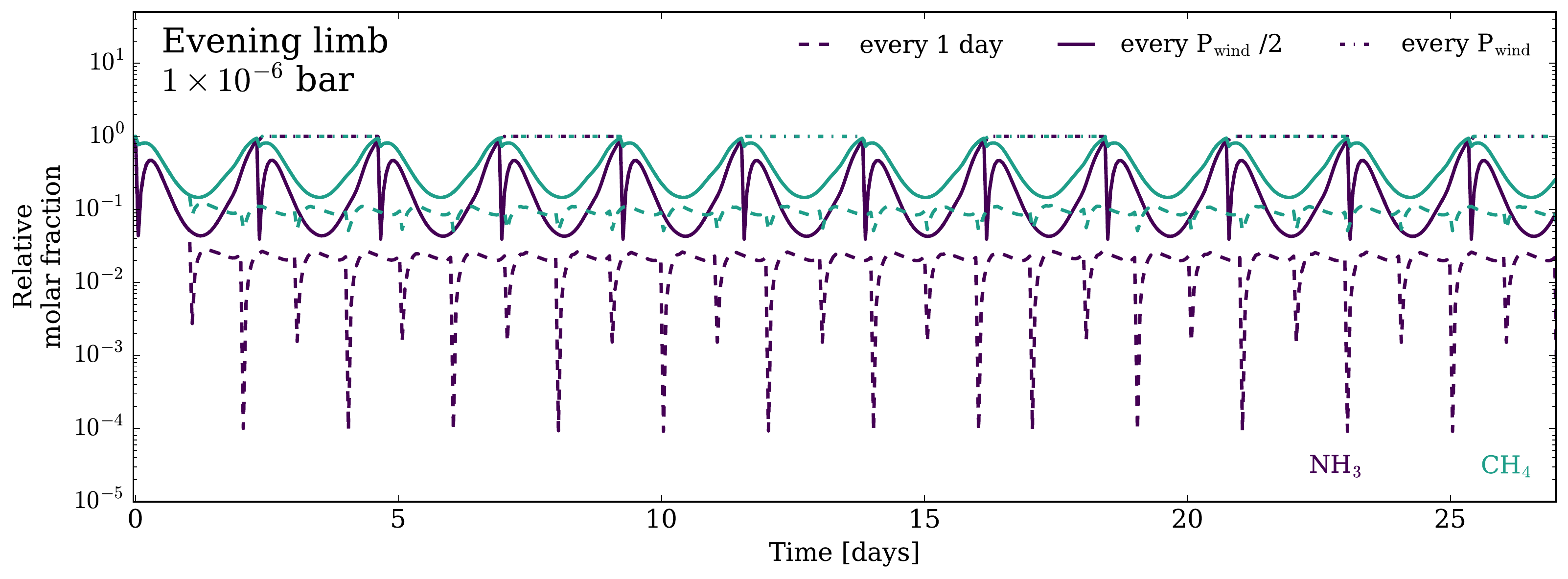}
	\includegraphics[width=17cm]{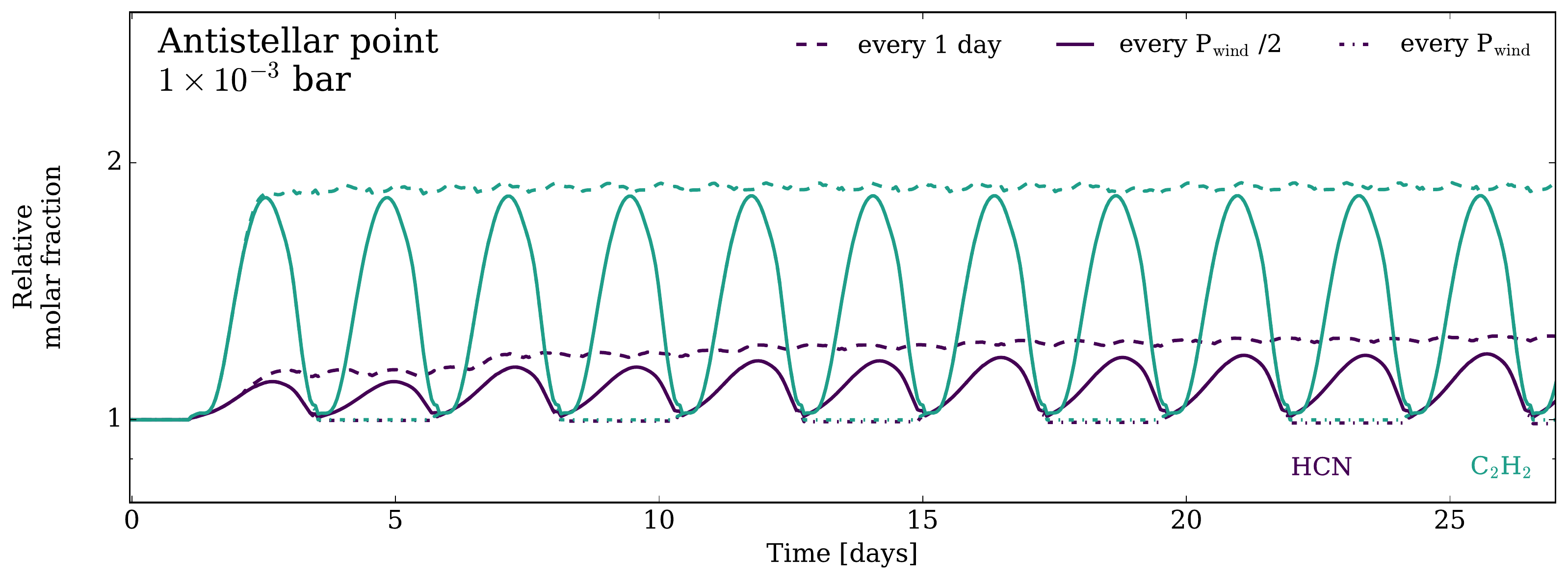}
	\caption{Evolution in time during the entire $\sim$1 month simulation span of $\methane$ and $\NHthree$ at $1 \mubar$ on the evening limb (\textit{upper panel}) and $\HCN$ and $\acetylene$ at $1 \mbar$ on the antistellar point (\textit{lower panel}). Three runs are shown where a $\num{2e33}$ erg flare event occurs every 1 day (\textit{dashed}), 2.3 days ($\Pwind/2$; \textit{solid}) and 4.6 days ($\Pwind$; \textit{dotted}). }
	\label{fig: results-Frequency-evolution}
\end{figure*}

Flares do not occur once in the lifetime of a star.
Instead, exoplanetary atmospheres will be flared repeatedly with a frequency that correlates with the total radiative output of the flare ($E_\mathrm{tot}$).
In this section, we explore the effects of repeated flaring in a simplified way by only considering $\num{2e33}$ erg flares, thereby omitting the inclusion of a FFD.
However, we consider the possible effects of such a FFD later in this paper (Section \ref{subsec: Results-ffd}).
For now, we consider flare periods of 1 day, 2.3 days ($\Pwind/2$) and 4.6 days ($\Pwind$).

Figure \ref{fig: results-Frequency-evolution} shows the evolution of $\CHfour$ and $\NHthree$ in the upper region ($1\mubar$) of the evening limb, and of $\HCN$ and $\acetylene$ in the middle layers ($1\mbar$) of the antistellar point.
%\textcolor{red}{HERE}
When the flare period equals 4.6 days, the atmosphere fully recovers to its pre-flare molar fraction between consecutive flare events.
This is, however, not the case when the flare period equals 1 day.
Instead, the continuously flared atmosphere maintains a perturbed state after 2-3 events, around which the molar fractions fluctuate.
For example, $\ammonia$ remains depleted by over an order of magnitude relative to its pre-flare value.
Although initially it takes several days to reach this new state, there does not seem to be a cumulative effect over longer timescales to which the atmosphere is continuously evolving towards.
The transition between these two regimes occurs when a flare period of 2.3 days is adopted.
In our model, it takes 2.3 days or $\Pwind/2$ to horizontally transport the entire flared dayside to the nightside.
In that same time interval, the unaffected nightside is now advected completely to the dayside.
Therefore, a new flare hitting the dayside after 2.3 days does not build on the effects of its precursor.
We nuance and discuss the validity of this result further in Sect. \ref{subsubsec: Discussion-superrotationAsAdvection}.
Finally, we echo that a $\num{2e33}$ erg flare on GJ 876 would occur roughly every 10 to 100 days (Section \ref{subsec: StellarFlares-flareEnergy}) and that a flare period of less than a week is unlikely.
However, we adopt the above flare periods to illustrate the boundary between lasting and non-lasting damage to the atmosphere, within the context of the adopted chemistry model.

\subsection{Impact of a flare frequency distribution (FFD)}
\label{subsec: Results-ffd}

\begin{figure*}
	\centering
	\includegraphics[width=17cm]{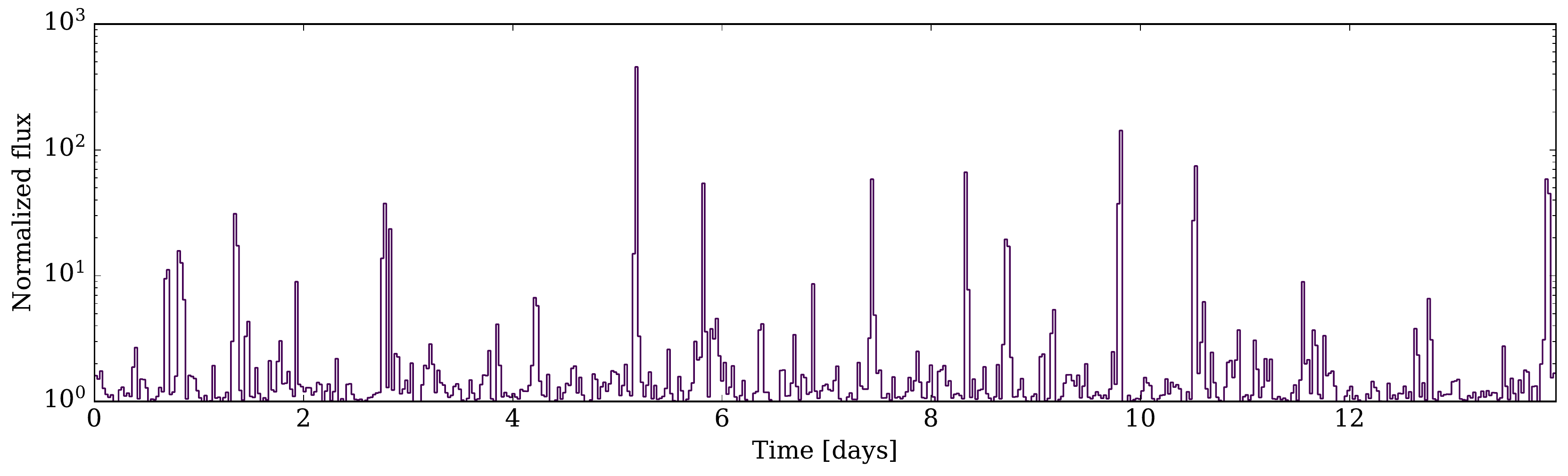}
	\caption{Light-curve of the adopted FFD in the Si IV doublet line (1393 \AA). The plotted flux is normalised to the quiescent value.}
	\label{fig: results-FFD-ffd}
\end{figure*}

Finally, we consider a more realistic case of repeated flaring by considering a range of flare energies and occurrences that adhere to a FFD.
We use the fiducial flare code of \citet{Loyd2018-TheMUSCLESSurveyV} to generate such a FFD that is shown in Fig. \ref{fig: results-FFD-ffd}.
The constructed FFD lasts about 2 weeks and all individual flares have a duration of $\tcol$.

%FFD lightcurve

\begin{figure*}
	\centering
	\includegraphics[width=17cm]{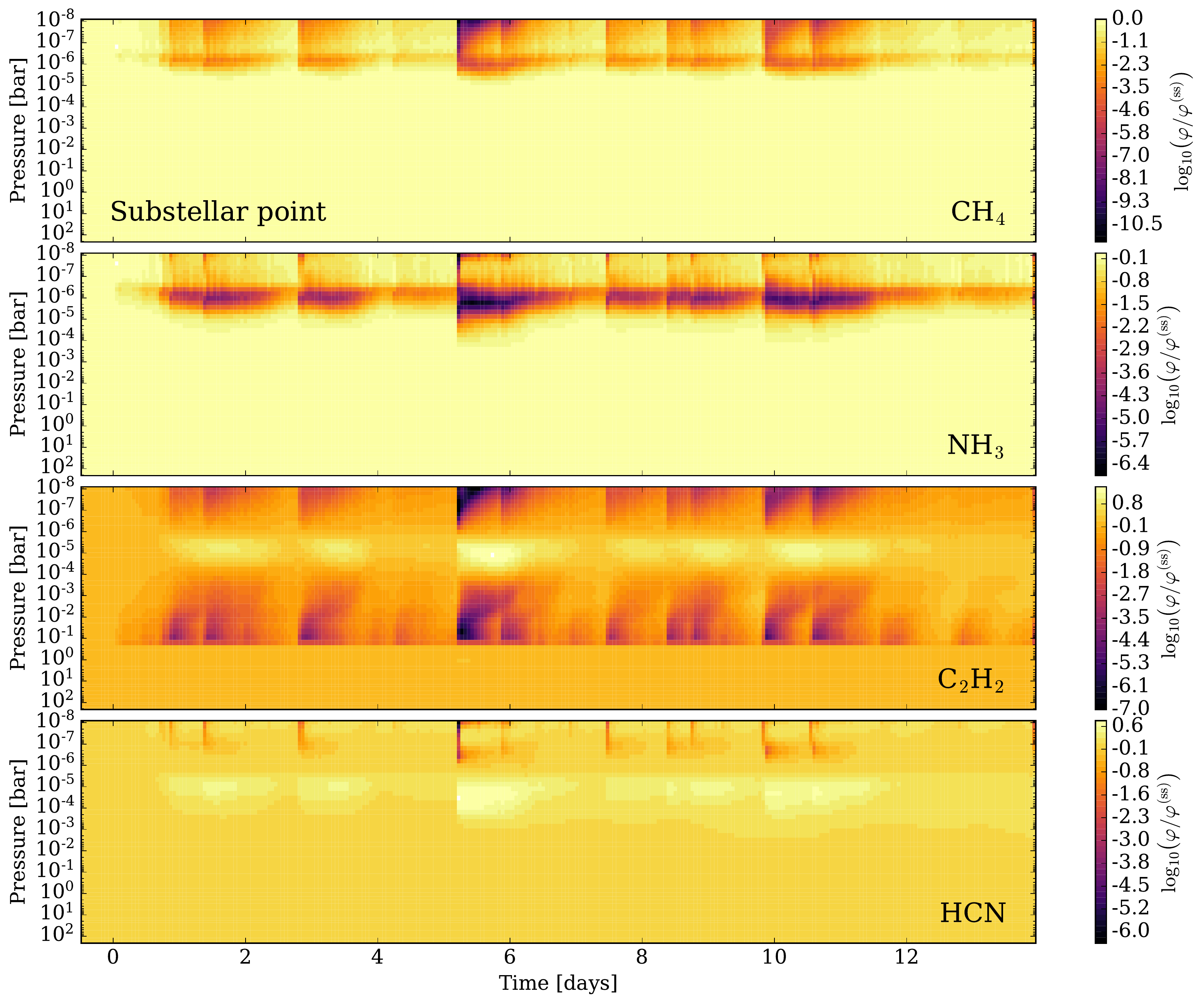}
	\caption{Evolution of $\CHfour$, $\NHthree$, $\acetylene$, and $\HCN$ throughout the vertical atmosphere on the substellar point during $\sim$two weeks of continuous flaring according to a FFD. The colourbar denotes the logarithm of the ratio between the current ($\varphi$) and the pre-flare, steady-state composition ($\varphi^{(\mathrm{ss})}$).}
	\label{fig: results-FFD-2D_plots_sub}
\end{figure*}

\begin{figure*}
	\centering
	\includegraphics[width=17cm]{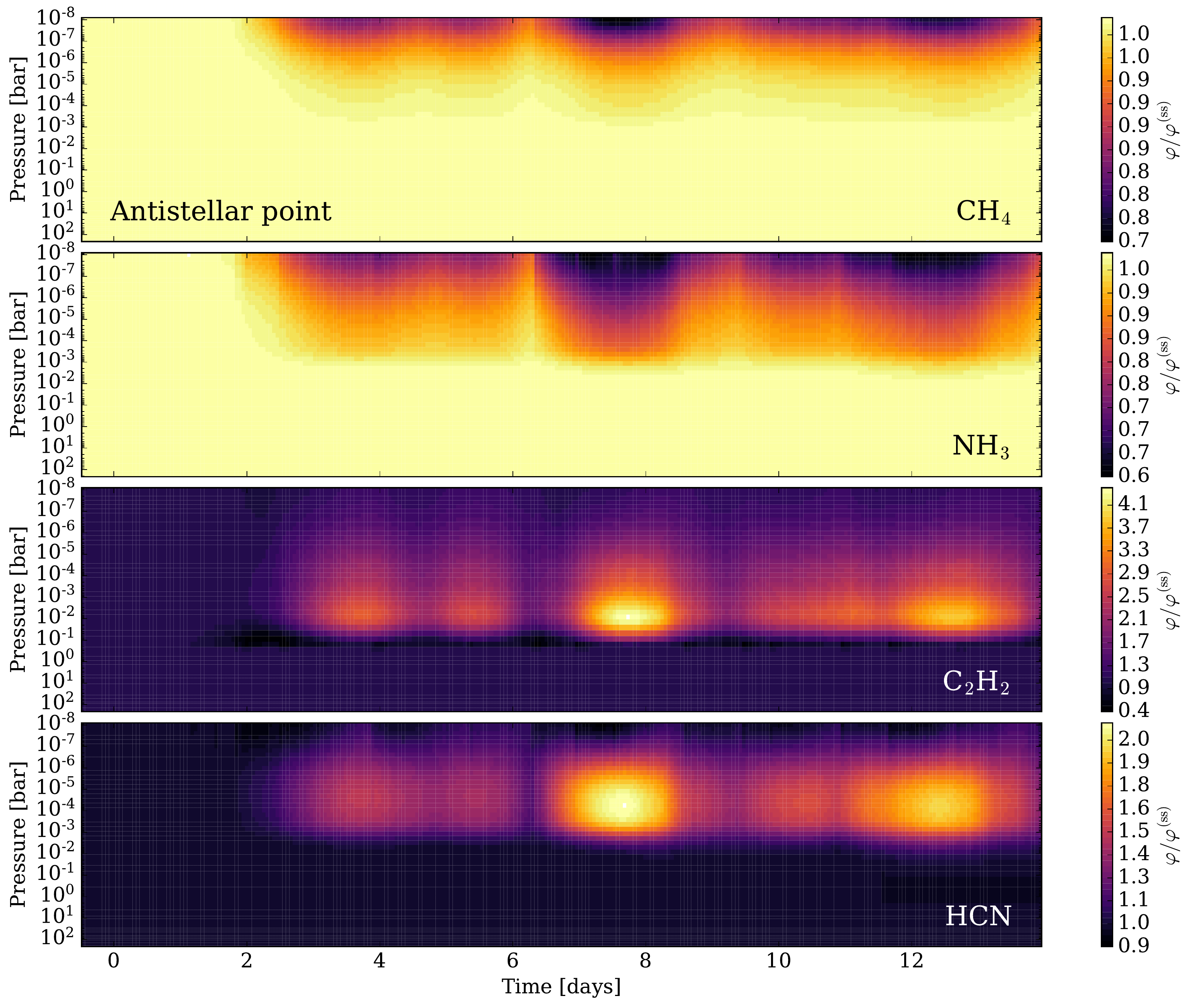}
	\caption{Same as Fig. \ref{fig: results-FFD-2D_plots_sub}, but for the antistellar point. Mind the different units for the colourbar.}
	\label{fig: results-FFD-2D_plots_anti}
\end{figure*}

Figures \ref{fig: results-FFD-2D_plots_sub} and \ref{fig: results-FFD-2D_plots_anti} show the variation in molar fraction compared to the pre-flare composition for $\CHfour$, $\NHthree$, $\acetylene$, and $\HCN$ on the substellar and antistellar point respectively. 
The results are in line with what we expect from the previous simulations, namely that mainly the upper atmosphere of the dayside is affected and the nightside become slightly enhanced with $\acetylene$ and $\HCN$.
Interestingly, the combination of high- and low-energy flares seems to keep the atmosphere in a composition that differs from pre-flare values, although there are clear fluctuations around this post-flare state.
To quantify these fluctuations, we can compute the median values for the abundances of certain molecules for different points in the atmosphere.
At $1 \mubar$ on the substellar point, the median molar fraction of $\CHfour$ is about a factor five lower than before the flaring and a factor $\sim$15 for $\NHthree$.
We show the evolution of other species in Appendix \ref{app: Other species during the ffd event}.
%2D plots

\begin{figure*}
	\centering
	\includegraphics[width=17cm]{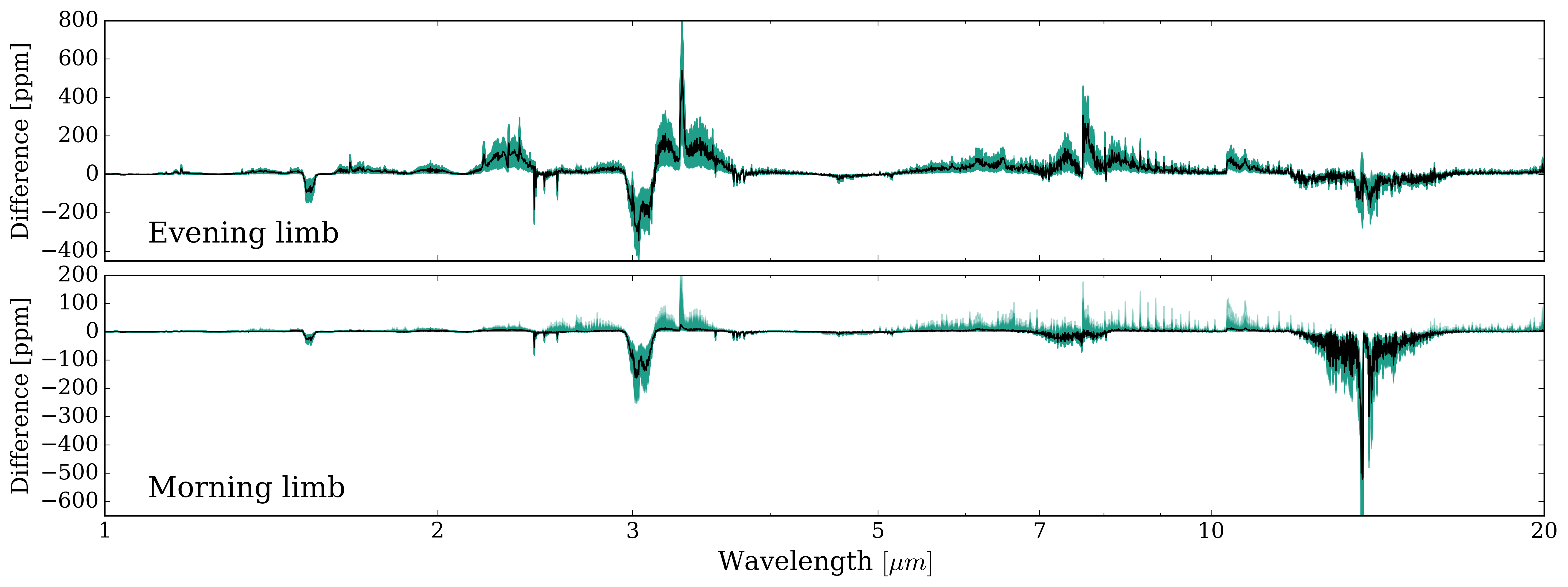}
	\caption{Differences in transmission spectra with respect to the pre-flare state on every time-step during the $\sim$two weeks of flaring according to a FFD, superimposed onto each other in green. The median value of the transit depth on every wavelength is indicated in black.}
	\label{fig: results-FFD-Transmission}
\end{figure*}

The FFD clearly manages to drive the atmospheric composition away from its pre-flare, steady-state distribution.
As shown in Sect. \ref{subsec: Results-transmission spectra}, we expect the perturbed composition to affect the transmission spectrum on different wavelengths, mainly due to $\CHfour$ depletion on the evening limb and $\acetylene$ production on the morning limb.
Figure \ref{fig: results-FFD-Transmission} shows the transmission spectra on every time-step during the simulation with FFD flaring, superimposed onto each other.
At several wavelengths, the transit depths do not return to the pre-flare values during the simulation.
Around $2.2 \um$, $3.3 \um$, and $7.8 \um$, $\CHfour$ decreases the median transit depth by $100$ to $200$ ppm on the evening limb, while $\acetylene$ increases the median value up to $500$ ppm at $3 \um$ and $14 \um$ on the morning limb.
As will be discussed in Sect. \ref{subsec: Discussion-observationalNoise?}, this result has important implications for future observations of similar close-orbiting planets around active stars.

%Transmission spectra

\section{Discussion} \label{sec: discussion}
\label{sec: Discussion}

\subsection{Comparison to previous work}
\label{subsec: Discussion-PreviousWork}

\subsubsection{\citet{Venot2016-Influenceof}}

As mentioned in Sect. \ref{sec:intro}, \citet{Venot2016-Influenceof} (V16 hereafter) have studied the consequences of a stellar flare on a sub-Neptune ($\Teff =1303$ K) and super-Earth ($\Teff =412$ K).
To describe the flare event, they used a time series based on the GFE of AD Leo that was constructed from observations by \citet{Segura2010-TheEffectofaStrong}.
The synthetic flare event used in this work contains a similar radiative output, relative to the quiescent flux levels, so that both equivalent durations are comparable (Section \ref{subsec: StellarFlares-flareEnergy}).
Therefore, the results of V16 can best be compared to the outcomes of our models where $E_\mathrm{tot}  \simeq \num{2e33}$ erg.
Additionally, V16 used a one-dimensional photo- and thermochemical model to study the neutral composition, in which they use the chemical network of \citet{Venot2012-Achemicalmodel}.
Aside from differences in thermal background structure, vertical mixing coefficients, planetary and stellar input radii/masses, etc., we model the longitudinal dimension with \citet{Agundez2014-Pseudo2Dchemicalmodel}'s pseudo-2D model and include the updated chemical network of \citet{Venot2020-Newchemicalscheme}.
Taking into account these differences between our work and the study of V16, only a qualitative comparison of the atmosphere's response can be drawn.

V16 found that the upper layers (< $1 \mubar$) of the super-Earth are depleted in ammonia by up to a factor $\sim$20 during the flare event, while the sub-Neptune becomes enriched by a factor $\sim$15.
For none of the different climates we considered (Section \ref{subsec: Results-different climates}), do we find any compelling production of ammonia.
%Only in the case of a XUV-flare (Section \ref{subsec: Results-flare wavelength}), we find a slight production in the upper layers on the dayside of our $\Teff =800$ K planet.
Furthermore, we find that the initial depletion of $\NHthree$ on the dayside during the flare event ranges from a factor 200 ($\textsc{run 0}$, Sect. \ref{subsec: Results-different climates}) to 1000 ($\textsc{run 2}$, Sect. \ref{subsec: Results-different climates}).
%ranges from only a factor two ($E_\mathrm{tot} =\num{2e32}$ erg) to over three orders of magnitude ($E_\mathrm{tot} =\num{2e33}$ erg).
We thus generally find a higher ammonia depletion by at least on order of magnitude, compared to V16.
%Given that the radiative output of AD Leo's GFE lies in between these values of $E_\mathrm{tot}$ for the star GJ 876 (when translated to equivalent duration $\delta$), the results of V16 fit within these ranges.
For example, we note that our pre-flare compositions differ from V16 as both their planets are subjected to high quiescent (X)UV-flux from the extremely active AD Leo, which already deprives the upper layers from photochemically reactive species such as ammonia.
Although we have focused mostly on four photochemically active molecules in this work, other species that are generally less affected such as $\CO$, $\COtwo$, $\NHtwo$, H, and $\OH$ are qualitatively impacted in similar ways (on the dayside) as in the models of V16.

V16 also addressed the implications on the transmission spectrum of the planet and found that mainly the $\COtwo$ features at $4.6$ and $14$ $\um$ are affected during the flare event.
This is in stark contrast to what we find, namely that the $\methane$ and $\acetylene$ opacities impact the morning and evening limb during the flare event.
A possible explanation for this is the discrepancy in pre-flare spectra, as this determines the relative sensitivity of the transit depths during a perturbation such as a flare event.
V16 found pronounced $\COtwo$ features in the pre-flare transmission spectra of both their models, while we do not reproduce this.
The prevailing spectral feature of $\COtwo$ is likely a direct result of the high average molar fraction ($\sim$$\num{e-4}$) that they compute in their pre-flare composition.
The pre-flare chemical state that we compute in this work and in B22 is much less rich in $\COtwo$ ($\varphi\simeq\num{e-7}$).
We note that V16 did not include $\acetylene$ in their radiative transfer calculations (only $\water$, $\COtwo$, $\OH$, $\HCN$, $\methane$, and $\CO$).

Finally, V16 also found that a single flare event can force the atmosphere to a new, post-flare chemical composition that converges to a new steady-state after $\sim$$\num{30 000}$ years.
This renewed state of the atmosphere then differs in the transmission spectra around $4.6$ and $14$ $\um$ (again $\COtwo$-features) by $\sim$40 ppm and $\sim$500 ppm for the super-Earth and sub-Neptune, respectively.
The value for the latter planet can reach up to $1200$ ppm when the atmosphere is flared every $\sim$5 hours.
V16 explained this phenomenon by hypothesising that a flare permanently modifies the photochemical rates so that the upper atmosphere is effectively bleached by the flare.
They further suggested that the lower layers maintain this new level of transparency long after the star returns to quiescence, which exposes the underlying layers to more intense radiation.
Although we are not able to extend our models to $\sim$30 000 years due to the linearity between physical and computation time (see Eq. \ref{eq: tcol-ColumnIntegrationTime}), we do not expect to find a post-flare composition that differs substantially from the pre-flare state.
By considering horizontal transport in our work, we effectively introduce an additional degree of freedom that is not present in the models of V16.
The fact that all of our single flare models have fully recovered their pre-flare composition by the end of our simulation time-domain ($\sim$1 month), and even recover within a factor five after one week, is in line with what we would expect from a realistic atmosphere when no other chemical sources or sinks are present compared to the pre-flare conditions.
A quantitative comparison and chemical pathway analysis between two models could further clarify the discrepancy.

%we strongly doubt that the post-flare state found by V16 is a physical consequence of the flare event.

%\begin{itemize}
%	\item Venot+16
%	\begin{itemize}
%		\item \citep{Venot2016-Influenceof} found a different post-flare steady-state composition with a signal in transmission that is significantly different from the pre-flare signal.
%		\item Their study was performed in 1D and therefore neglected the important role of circulation dynamics in redistributing chemical compounds.
%		\item (We cannot simulate 33 000 years due to linearity of our models)
%%		\item Based on findings of B21, B22, (other dynamics studies), we anticipate to find no such significant difference between the pre- and post-flare state.
%		\item Arguments why: (1) Numerically, initial perturbation by flare is a direct change of boundary conditions. Once the SED returns to quiescence
%	\end{itemize}
%\end{itemize}

\subsubsection{\citet{Louca2022+Theimpactof}}
%\subsubsection{TODO \citet{Louca2022+Theimpactof}}

\citet{Louca2022+Theimpactof} (L22 hereafter) considered three $\Htwo$-dominated planet atmospheres of GJ 876c, GJ 832c, and GJ581c, with host stars that have been observed in the MUSCLES survey \citep{France2016-TheMUSCLESTreasurySurveyIMotivationandOverview}.
They subsequently used the fiducial flare code of \citet{Loyd2018-TheMUSCLESSurveyV} to generate FFDs and model the one-dimensional atmospheric composition.
They find that the abundances of several species including $\methane$, $\COtwo$, and $\water$ do not recover to pre-flare values in between subsequent flares and some even accumulate gradually over time.
Based on the result presented in Sect. \ref{subsec: Results-ffd}, we also find that several molecules (e.g. H, $\CHfour$) stay in a perturbed state with median molar fractions differing from the pre-flare values.
However, for none of the species considered in this work, we see a clear trend that points to the accumulation in time.
It can be argued that the horizontal transport prevents the atmosphere from accumulating flare effects by continuously replenishing the dayside with material from the nightside.
%In particular, atomic hydrogen achieves an abundance of nearly a factor 10 increase compared to the pre-flare value in between 1 bar and 10 $\mbar$.
Furthermore, L22 do not find that the changes in chemical composition affect the transmission spectra by more than several ppm, which is in contrast to what we find (see Fig. \ref{fig: results-FFD-Transmission}).
%We note that L22 do not take into account $\acetylene$ during their radiative transfer calculations.
%Nevertheless, also the upper atmospheric depletion of $\CHfour$ is not sufficient to drastically affect their transmission spectra.
%Finally, they speculate that the gradual incline of $\COtwo$ due to repeated flaring could result in JWST-observable changes in the transmission spectrum, although this statement is highly speculative as it requires linear extrapolation over long periods in time.
Just like the discrepancy with V16, a more detailed comparison of the pre-flare conditions and model set-up is needed to further investigate these differences.

\subsection{Model improvements and future work}

%We have not accounted for several processes, that might impact our results.

%\subsubsection{Photoevaporation}
%All forms of thermal and non-thermal escape processes have been neglected in this work.
%
%Several studies have explored the possibility that stellar flares (and accompanying particle events, see later) introduce variability to the atmospheric photoevaporation of exoplanets \citep{Chadney2017-Effectofstellarflares, Atri2017-Modellingstellarproton, Hazra2022-TheimpactofCMEs}, which could explain observations of fluctuating Ly $\alpha$ transits \citep{LecavelierdesEtangs2012-Temporalvariations, Bourrier2013AtmosphericEscapefromHD189733b}.
%
%
%Mainly because the effect is limited on a one month period.
%
%We have not considered the effects of photo-evaporation. See Hazra+2021 -- photo-evaporation during CME and flares on HD189733b. This is especially relevant for the XUV flare (in addition with ionization)

\subsubsection{Stellar particle events}
\label{subsubsec: Discussion-SPE}

Stellar flares are often accompanied by SPEs that eject (charged) energetic particles into space.
Such SPEs can impact a planetary atmosphere by inducing additional photochemistry \citep{Segura2010-TheEffectofaStrong}, enhancing ionization levels \citep{Barth2021--MOVESIV}, and increasing mass-loss rates \citep{Chadney2017-Effectofstellarflares}.
Because chemical models of planetary atmospheres that undergo SPEs have mainly been constructed to investigate the consequences on habitability \citep[e.g.][]{Segura2010-TheEffectofaStrong,Tabataba-Vakili2016-AtmosphericEffectsOfStellarC, Tilley2019-ModelingRepeatedMDwarf, Atri2020-StellarEnergeticParticles}, little attention has been given to close-orbiting gaseous planets.
A follow-up study, that assesses the combined effects of flares and SPEs on the chemical composition of such gas giants with a multi-dimensional kinetics model, would therefore be valuable.

%Aside from stellar flares and SPEs, active stars have stellar winds and can abruptly shed material as so-called Coronal Mass Ejections.

\subsubsection{Temperature structure}
\label{subsubsec: Discussion-Temperature}

Throughout the chemistry simulations, we have fixed the thermal background to the two-dimensional distribution as was calculated for steady-state conditions by \citet{Baeyens2021-GridofPseudo2D}.
However, the temperature distribution might be affected during a flare event by the sudden increase of irradiation, change in chemical composition and even accompanying SPE.
Although \citet{Segura2010-TheEffectofaStrong} found that the temperature profile is perturbed by only a few Kelvin during and after a flare event of this magnitude, \citet{Venot2016-Influenceof} correctly argued that the radiative response of an atmosphere to a stellar flare will be much different for warmer, gaseous planets.
More research is needed to properly asses the impact of a flare on the temperature structure of, for instance, an 800 K planet such as considered in this work.

%, one should couple a flare event to a general circulation model (GCM) with self-consistent radiate transfer, and potentially even include disequilibrium chemistry via chemical kinetics (such as presented by \citet{Drummond2020-Implicationsofthreed}), with photochemistry. 

\subsubsection{Clouds and haze}
\label{subsubsec: Discussion-CloudsAndHaze}

We have neglected clouds and haze in our chemistry simulations and calculations of transmission spectra to isolate flare-induced changes to gas-phase molecules.
However, aerosols are thought to be present in the bulk of exoplanet atmospheres \citep[e.g.][]{Marley2013-CloudsandHazes, Gao2021-Aerosolsinexo}.
Therefore, we must place the findings presented in this work in the context of cloudy and hazy atmospheres.

Aerosols are likely to play an important role in the heat distribution of planets by back-warming the atmospheric layers below the cloud deck (greenhouse effect), and by increasing the albedo and thereby changing the incoming radiative energy \citep{Helling2019-ExoplanetClouds, Gao2021-Aerosolsinexo}.
The question then arises whether clouds can effectively lower photolysis rates on exoplanets by shielding the atmosphere from radiation, as has been shown for Earth's atmosphere \citep{Hall2018-cloudImpactsonPhotochemistry}.
Although it has not been modelled for exoplanet atmospheres, this effect could moderate the flare-impact on the molar fractions of photochemically active species.

Depending on the altitude of the aerosols, clouds and haze can also be responsible for flattening molecular absorption features in the transmission spectra of planets \citep[e.g.][]{Pont2008-Detectionofatmospherichaze, Bean2010-Agroundbasedtransmission, Kreidberg2014-Cloudsintheatmosphere}.
Although aerosols mainly affect the optical regime through Rayleigh scattering, the entire wavelength range can be affected depending on the particle size, composition and altitude of the aerosols \citep{Gao2021-Aerosolsinexo}.
This could have implications on the values for flare-induced fluctuations in the transmission spectrum, reported in this work.

Finally, we have found that flares can increase molar fractions of hydro-carbons when photodissociated species from the dayside enter the nightside through horizontal transport.
In particular, $\acetylene$ is a precursor of more complex hydrocarbons and leads to haze formation.
It is unclear how such a short-term perturbation to the gas-phase chemistry affects haze formation and whether this may introduce additional variations to the (optical) transit depths.

%\{strict dependence of our results on $\tauzonal$}: Although $\tauzonal$ is typically computed as a timescale for zonal mixing (Drummond, B21), it's role - as a exact quantity that determines the advection period of wind parcels - plays a crucial role in explaining our results.
%
%We want to properly nuance our results with this realisation. 
%For example, in reality, we do not expect an exact differentiation between a/no cumulative effect of repeated flaring at $\tauzonal$. This results should much rather also be considered as a timescale argument.

\subsubsection{Three-dimensional circulation}
\label{subsubsec: Discussion-3D circulation}
\begin{figure}
	\resizebox{\hsize}{!}{\includegraphics{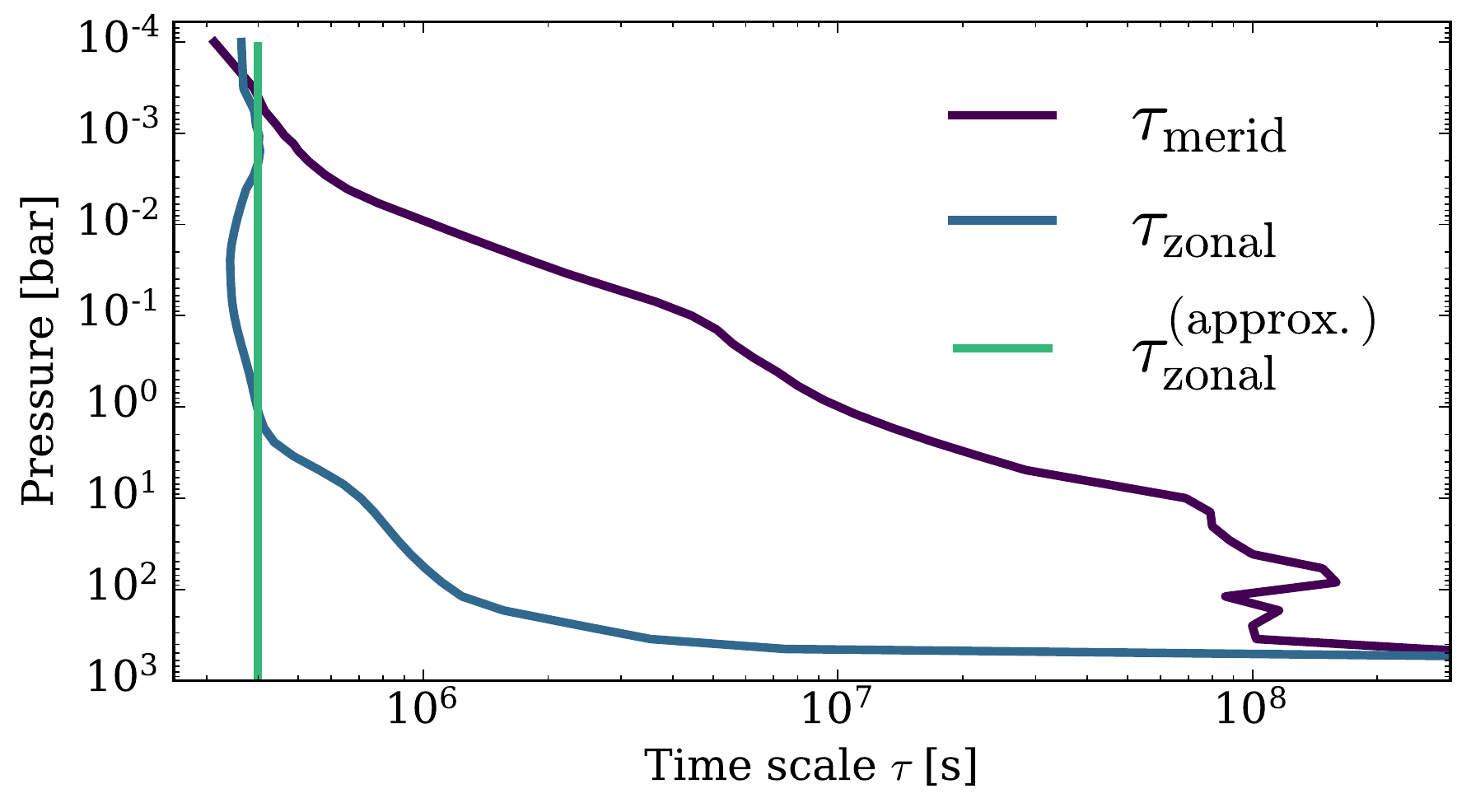}}
	\caption{Zonal ($\tau_\mathrm{zonal}$) and meridional ($\tau_\mathrm{merid}$) advection timescale computed from the GCM outcome of the model planet corresponding to $\Teff =800$ K with a M dwarf host star, presented in \citet{Baeyens2021-GridofPseudo2D}. An approximate zonal timescale ($\tau^{(\mathrm{approx.})}_\mathrm{zonal}$ = $\Pwind$) is computed with the fixed value for $\vwind$ used the chemistry simulations.}
	\label{fig: discussion-timescales}
\end{figure}

The climates of tidally locked, close-orbiting gaseous planets are three-dimensional, and aside from equatorial superrotation, their circulation pattern consists of a variety of flows.
The equatorial middle layers ($\sim$1 bar to $\sim$1$\mbar$) develop strong prograde winds, while a combination of pro- and retrograde flows is typically present at higher altitudes ($\lesssim 0.1\mbar$).
The upper boundary of general circulation models typically ranges between $\num{0.1} \mbar$ and $\num{1} \mubar$, to avoid numerical issues.
Therefore, we are limited in information about the layers above these pressure values, which are generally of most interest for photochemistry. 
%Early circulation models of hot Jupiters, coupled to a Newtonian radiative scheme typically found peak values for the horizontal wind speeds around the $10$ to $100 \mbar$ regime \citep{Cooper2005-DynamicMeteorology, CooperShowman2006-DynamicsandDisequilibrium, Rauscher2010-ThreedimensionalModelingofHotJupiterAtmosphericFlows, Mayne2014-Theunifiedmodelafull}.
%However, studies that adopted more accurate radiation schemes revealed that equatorial superrotation does not or only very slightly diminishes in strength near the $\num{0.1} \mbar$ boundary \citep{Showman2009-AtmosphericCirculationofHotJupiters, Amundsen2016-TheUKMetOffice, Deitrick2020-THOR2.0MajorImprovements, Drummond2020-Implicationsofthreed, Schneider2022-Exploringthedeepatmospheres}.
Some three-dimensional models of hot Jupiter exhibit jet streams that diminish in speed near the upper atmosphere \citep{Cooper2005-DynamicMeteorology, CooperShowman2006-DynamicsandDisequilibrium, Rauscher2010-ThreedimensionalModelingofHotJupiterAtmosphericFlows, Mayne2014-Theunifiedmodelafull}, whereas other models sustain a superrotating jet stream up to the upper boundary of the simulation domain \citep{Showman2009-AtmosphericCirculationofHotJupiters, Amundsen2016-TheUKMetOffice, Deitrick2020-THOR2.0MajorImprovements, Drummond2020-Implicationsofthreed, Schneider2022-Exploringthedeepatmospheres}. 
Possible reasons for this discrepancy may be the radiative scheme \citep[see discussion in Appendix D of][]{Schneider2022-Exploringthedeepatmospheres} or the upper boundary treatment.
In this work, we have assumed that horizontal winds dominate the entire equatorial band, including the upper layers down to $\num{e-8}$ bar.
To quantify the validity of this assumption, we use timescale arguments based on the GCM outcome of the model planet corresponding to $\Teff =800$ K with a M dwarf host star, presented in \citet{Baeyens2021-GridofPseudo2D}.
In particular, we compute the meridional advection timescale $\tau_\mathrm{merid} \simeq \pi\rplanet / 2 \rvert v \rvert$ with $v$ the meridional wind speeds and the zonal advection timescale $\tau_\mathrm{zonal} \simeq 2\pi\rplanet / \rvert u \rvert $ with $u$ the zonal wind speeds \citep[][]{Drummond2020-Implicationsofthreed}.
Both values are averaged over all longitudes and the equatorial latitudes ($-20\degrees < \phi < 20\degrees$) and are shown in Fig. \ref{fig: discussion-timescales}.
The fact that $\tau_\mathrm{zonal} \ll \tau_\mathrm{merid}$ for pressures above $\sim$$0.1 \mbar$ illustrates that our assumption, that horizontal winds dominate the circulation pattern, is valid for the middle and deep atmosphere (down to $\sim$$0.1 \mbar$).
Only in the uppermost regions does meridional advection become equally or more important than horizontal transport.
However, the flare impacts these upper layers the most (e.g. Fig. \ref{fig: results-chemComp-2.3daysevolutionOfLineProfiles}) and it is possible that the inclusion of more detailed flows will impact our findings considerably.
%However note that in the atmospheric layers below, $\tau_\mathrm{zonal}$ exceeds $\tau_\mathrm{merid}$  by several orders of magnitude.

At lower and higher latitudes ($\rvert \phi \rvert \gtrsim 50 \degrees$) there can even be direct day-to-night flows over the poles at high altitudes.
Both the equatorial meridional advection and polar day-to-night flows at high altitudes are likely to also impact the atmosphere's recovery after a flare event by transporting and mixing material away from the flared dayside.

\subsubsection{Superrotation as advection}
\label{subsubsec: Discussion-superrotationAsAdvection}

The premise of the pseudo-2D chemistry model of \citet{Agundez2014-Pseudo2Dchemicalmodel} is that equatorial superrotation can be approximated by a uniform zonal wind that acts as advection on the entire atmospheric column (solid-body rotation).
Aside from reducing the complex three-dimensional circulation pattern to the dominating jet stream (see Sect. \ref{subsubsec: Discussion-3D circulation}), the assumption of solid-body rotation with a fixed period $\Pwind$ ($\sim$4.6 days) is clearly a determining factor in our models for the evolution of the atmospheric composition after the flare event (e.g. Figs. \ref{fig: results-chemComp-2.3daysevolutionOfLineProfiles}, \ref{fig: results-chemComp-C2H2_evolution}, \ref{fig: results-climates_run01}).
We found that the evening limb is perturbed in the first $\sim$2.3 days ($\Pwind/2$) after the start of the flare, while the morning limb remains unaffected several hours after the event.
Between $\sim$2.3 to $\sim$4.6 days, the morning limb becomes affected while the evening limb fully recovers. 
However, this abrupt transition at $\Pwind/2$ is an artefact of the chemistry model, and we do not expect such a time evolution in reality.
Instead, additional processes such as small-scale turbulence and non-zonal mixing (see Sect. \ref{subsubsec: Discussion-3D circulation}), will initiate further mixing of atmospheric layers.
As such, we expect diffusion to modulate the patterns that we obtain from our purely advective models.
Therefore, one should consider this transition value of $\Pwind/2$ as an estimate, rather than a exact quantity.
The same nuance holds for the results presented in Sect. \ref{subsec: Results-repeated flaring}, where we found that repeated flaring only manages to permanently perturb the atmosphere to a post-flare state when the flare period equals $\Pwind/2$ or less.

\subsubsection{Three-dimensional transmission spectroscopy}
\label{subsubsec: Discussion-3D spectra}

The transmission spectra presented in this work are computed with a thermal structure and molecular composition representative for the equator.
However, transit spectroscopy probes the full terminator of the projected planetary disk, so the information from higher latitudes is missing from our models.
Nevertheless, the equatorial region, which is generally hotter than the polar regions, is expected to dominate the spectral signature \citep[e.g.][]{Caldas2019-Effectsofafully3D, Baeyens2021-GridofPseudo2D}.

%Additionally, 
%we are currently not yet able to probe the ingress and egress of a transiting exoplanets separately.
%Therefore,
Additionally, in this work, we focussed on flare-induced variations to the morning and evening limb separately, while both would contribute to the transmission spectra.
The values reported in this work for flare-induced fluctuations on individual terminators are therefore likely upper limits as actual observations will be an average between the evening and morning limb contributions.
We note that recently the ingress and egress of transiting exoplanets have been probed separately \citep{Ehrenreich2020-Nightsidecondensation, Espinoza2021-ConstrainingMorningsandEvenings}, which provide unique insights on the three-dimensional structure of exoplanets.
Finally, we note that recent studies which have researched asymmetries in three-dimensional transmission spectra indicate that the vertically extended evening limb contributes more than the less extended morning limb \citep{Caldas2019-Effectsofafully3D, Falco2021-Towardamultidimensional-I, Pluriel2021-Towardamultidimensional-II, Lee2021-3DradiativeTransferfor, MacDonald2021-TRIDENT, Nixon2022-AURA3D}.

\subsection{Implications for observations}
\subsubsection{Additional uncertainty to JWST data}
\label{subsec: Discussion-observationalNoise?}

Considering our finding that repeated flaring according to a FFD can induce > 50-100 ppm variations in the transmission spectra of the modelled close-orbiting gaseous planet(s), one can wonder what the implications are for the James Webb Space Telescope (JWST) that is estimated to achieve a similar precision \citep{Barstow2015-Transitspectroscopy, Greene2016-CharacterizingTransiting}.
In particular, repeated flaring could introduce an additional component of uncertainty to JWST data and may even result in measurable variability.
Indeed, Fig. \ref{fig: results-FFD-Transmission} indicated that the $3.3 \um$ $\CHfour$-feature and $14 \um$ $\acetylene$-feature experience fluctuations of 100 to 300 ppm around a median transit depth.
Such fluctuations could introduce an uncertainty to the observations of JWST, which in turn propagate into uncertainty on the retrieved abundances of $\CHfour$, $\acetylene$, and potentially other species as well.
Furthermore, the median transit depths of the before mentioned spectral features also differ by several hundreds of ppm from the values obtained when assuming stellar quiescence.
This indicates that, if one does not account for flaring of the active host star during the analysis, one might bias the retrieved abundances.
In order to quantify the above effects, one should perform retrievals on the computed transmission spectra, but this is out of scope for this work.
In order to account for such changes in the modelling of planets around active stars, we consider it beneficial to monitor the planet's host star in the UV up to several days before and during the scheduled planet observations.

In Sect. \ref{subsec: Results-repeated flaring}, we have also investigated repeated flaring for an extreme case of frequent high energy flares, and found that the atmosphere almost completely returns to its pre-flare state when the flare period is longer than a few days (i.e. $\Pwind/2$ or 2.3 days for the considered model).
The above conjecture heavily depends on $\Pwind$, and by extension $\tauzonal$, which varies among planets with different climates, but is generally on the order of several days. 
Although we have nuanced the importance of the $\Pwind$ parameter in our chemistry models in Sect. \ref{subsubsec: Discussion-superrotationAsAdvection}, and investigated the effects of repeated flaring for a more realistic case by means of a FFD (Section \ref{subsec: Results-ffd}), this result indicates that $\tauzonal$ can be a valuable parameter in assessing how efficient the atmosphere recovers in between subsequent flare events.

Finally, \citet{Baeyens2022-GridIIphotochemistry} have noted that photochemical processes have a limited impact on transit spectra in general, but high-resolution observations of exoplanet atmospheres are more likely to be affected by photochemistry.
Therefore, we stress that any transmission-spectroscopic effects of flaring discussed in this work are potentially even more pronounced in high-resolution spectra.

\subsubsection{High-cadence monitoring of active host stars}

%\{After examining a variety of flares, we identify that flares of different energy change the atmosphere's chemical response only quantitatively, and not qualitatively, and that longer duration flares seem to have a slightly larger impact due to their prolonged exponential decay.
%	Both the importance of total radiative output and time-evolution from peak to quiescence illustrates the need for high-cadence monitoring of active host stars.}

We have shown that both the flare's total outputted radiative energy (Section \ref{subsec: Results-flare energy}) and exponential decay from peak to quiescence (Section \ref{subsec: Results-flare duration}) are important factors that control the impact on the planetary atmosphere.
Current flare observations often have a poor time-sampling as a result of low-cadence monitoring, which can result in a degeneracy between sharply peaked and weakly peaked flares \citep{Howard2021-NoSuchThingAsaSimpleFlare}, thereby losing critical information.
Therefore, we advocate for high-cadence observations, such as the TESS 20-second cadence mode, for future monitoring of active host stars.

%
%Other stuff
%\begin{itemize}
%	\item LRS mode is not suitable for detecting the 14 um feature of C2H2
%	\item MIRI MRS mode is more suitable BUT... Read-out problems for transmissions spectroscopy?
%	\item MRS requires 3 different visits for one full coverage of the 4.9 - 28.3 um range.
%\end{itemize}

%Note that the conclusions presented in Section \ref{subsec: Results-chemical composition} and \ref{subsec: Results-transmission spectra} still hold.

%\subsection{Fiducial flare of \citet{Loyd2018-TheMUSCLESSurveyV}}
%As part of the MUSCLES survey (cite), \citet{Loyd2018-TheMUSCLESSurveyV} developed a small python routine that generates a fiducial UV flare for an M dwarf in an attempt to facilitate consistent model comparisons.
%We take ... and re-run our models with such flare prescription.

\section{Conclusions} \label{sec: conclusions}
We have studied the effect of stellar flares on the chemical composition of a $\Teff =800$ K, tidally locked gaseous exoplanet.
By combining a series of pseudo-2D chemistry models, we track the evolution of the two-dimensional molecular composition around the equatorial band, where the climate is dominated by a jet stream of $\sim$$1.5 \kms$ that advects an air parcel around the planet in $\sim$4.6 days ($\Pwind$).
Simultaneously, we compute one-dimensional transmission spectra on both the evening and morning limb to assess the impact of fluctuations in the chemical composition.
We have constructed a $\sim$$37$ min flare of $E_\mathrm{tot} =\num{2e33}$ erg from the SED of GJ 876 as a vantage point to explore flares that vary in total energy, duration, and occurrence frequency.
%\{We have constructed a flare of $E_\mathrm{tot} =\num{2e33}$ erg from the SED of GJ 876 that closely resembles the Great Flare Event of AD Leo \citep{Hawley1991-TheGreatFlare} as a vantage point to explore flares that vary in total energy, duration, and occurrence frequency.}
We find that photochemically active and abundant species such as $\CHfour$, $\NHthree$, $\acetylene$, and $\HCN$ are affected throughout the atmosphere.
Furthermore, we show that, after a single flare event, the evening limb remains in a perturbed state until $\sim$2.3 days ($\Pwind/2$) after the start of the flare event, as then the flared dayside is completely advected to the nightside.
Further advection from $\sim$2.3 to $\sim$4.6 days enhances the morning limb with $\acetylene$ and $\HCN$, which are produced by photodissociated ammonia and methane on the nightside \citep{Moses2014-Chemicalkinetics}.
We find that mainly methane depletion on the evening limb reduces transit depths by 100 to 300 ppm (t < $2.3$ days) and acetylene production on the morning limb increases the 14 $\um$ feature up to 300 ppm ($2.3$ days < t < $4.6$ days).
After $\sim$4.6 days, we see that $\acetylene$ is advected with slightly enhanced molar fractions around the planet for up to two weeks after the flare event, mainly in the deeper layers (e.g. at $1 \mbar$), although this affects the spectra by less than 20 ppm.
Because the three-dimensional circulation pattern of our hypothetical planet is simplified to a uniform zonal wind that advects an air parcel in exactly $\sim$$4.6$ days, we argue that the above timeline, and in particular the transition on $\sim$2.3 days, should be considered an estimate rather than an exact value.

A flared planet with effective temperature of 1600 K, that orbits closer to the star, experiences more depletion on the dayside in the upper atmosphere, is affected less deep in the vertical atmosphere due to shorter chemical timescales and recovers faster than the 800 K planet.
%After examining a variety of flares that differ in energy and duration, we identify that \{mainly the radiative energy contained at the start of the flare event} plays an important role in determining the atmosphere's response to the flare event, which illustrates the need for high-cadence monitoring of active host stars.
After examining a variety of flares, we identify that flares of different energy change the atmosphere's chemical response only quantitatively, and not qualitatively, and that longer duration flares seem to have a slightly larger impact due to their prolonged exponential decay.
Both the importance of total radiative output and time evolution from peak to quiescence illustrate the need for high-cadence monitoring of active host stars.
By considering multiple flare events following a FFD, we have shown that repeated flaring has the potential to permanently alter the chemical composition at several locations in the atmosphere.
Furthermore, depletion of methane on the evening limb and production of acetylene on the morning limb can permanently alter the transit depth at wavelengths such as $3.3$ and $14 \um$ by several hundreds of ppm, with fluctuations in the range from 100 to 300 ppm.
We argue that the above effects have important implications for observations that will be acquired with JWST.
Detectable variations in photochemically active molecules can be expected, and their retrieved abundances could be biased if flaring of the host star is not accounted for.
Therefore, we recommend that transmission studies with JWST are accompanied and even preceded by simultaneous (UV) observations of the host star to further clarify the exoplanet's atmospheric response to flaring.

%\{We argue that the above effects have important implications for observations that will be acquired with JWST, as retrieval models should include the effects of stellar flares in their treatment of disequilibrium chemistry to avoid biases on the abundances of several species.}

% if the flare period is equal to or less than $\sim$$2.3$ days.
%If the flare period is higher, horizontal transport of the flared dayside to the nightside is efficient enough so that the atmosphere recovers to an extent where a subsequent flare does not accumulate on the perturbation of its predecessor.

%Flare frequency distributions of M dwarfs suggest that flares with an integrated radiative output of $\sim\num{e32}$ erg or lower (relative to the quiescent luminosity of GJ 876) qualify to occur every few days, but we find only a limited impact for flares of this energy.

%However, given that $\sim\num{e32}$ erg flares can still alter the transmission spectra up to 150 ppm, we argue that frequent, low-energy flares can introduce uncertainty to the data obtained by state-of-the-art telescopes such as JWST.

%Future work that considers a range of flare energies by implementation of a more realistic flare frequency distribution would be beneficial.

\begin{acknowledgements}
	
The authors thank Marcelino Ag\'undez for the use of the pseudo-2D chemistry code and Parke Loyd for creating the open source fiducial flare model.
We also thank the referee for their insightful comments that highly improved this work.
TK, RB, and LD acknowledge support from the KU Leuven IDN grant IDN/19/028.
RB acknowledges funding from a PhD fellowship of the Research Foundation $-$ Flanders (FWO).
RB and LD acknowledge support from the FWO research grant G086217N.

\end{acknowledgements}

\bibliographystyle{aa}
\bibliography{stellarflares}

\begin{appendix}

\section{Evolution of constituents not discussed in Sect. \ref{subsec: Results-chemical composition}}
\label{app: Other species during the single flare event}

\begin{figure*}
	\centering
	\includegraphics[width=17cm]{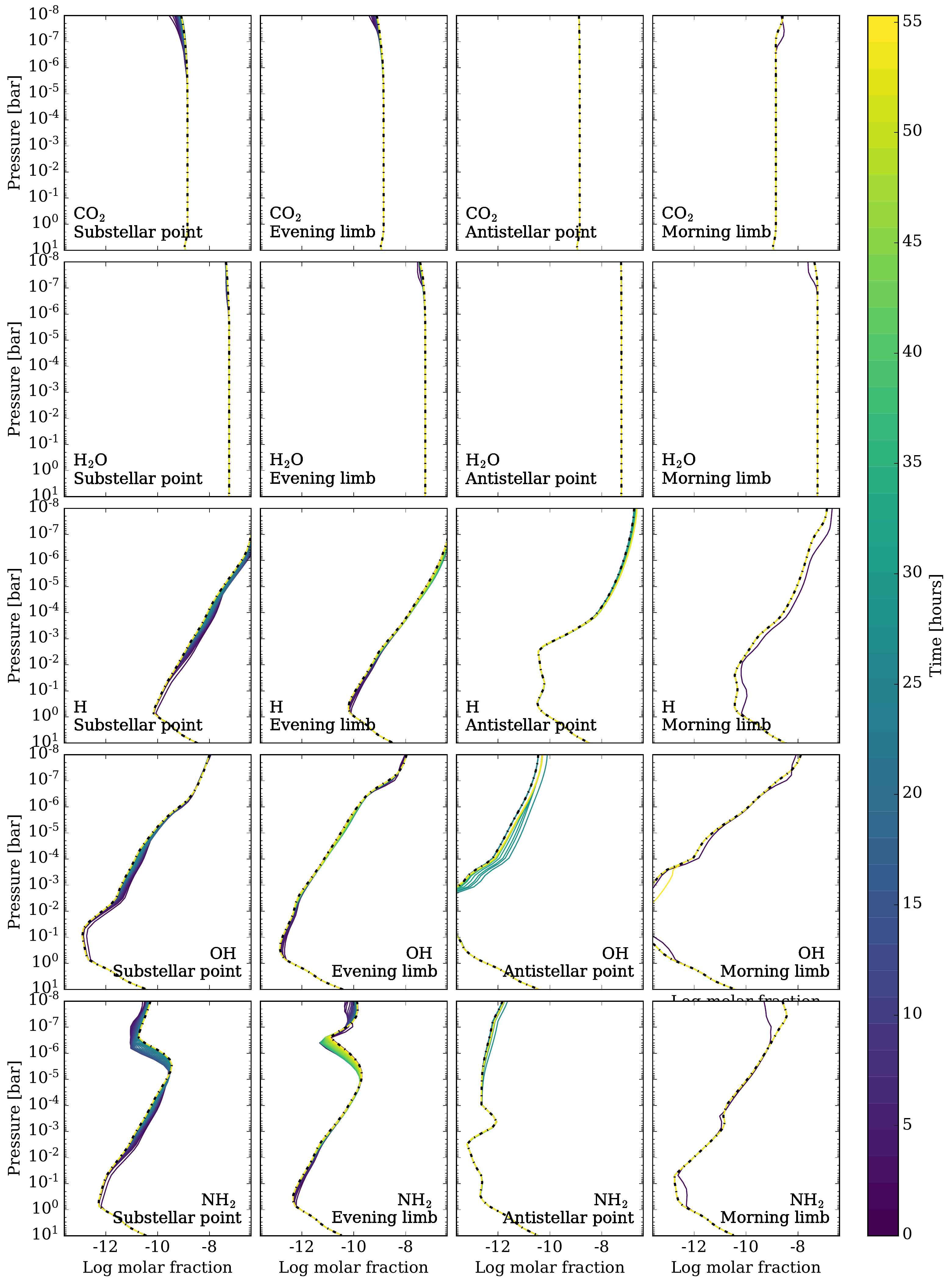}
	\caption{Same as Fig. \ref{fig: results-chemComp-2.3daysevolutionOfLineProfiles}, but for  $\COtwo$, $\water$, H, $\OH$, and $\NHtwo$. We note that the range on the horizontal axis is now extended to lower molar fractions.}
	\label{fig: appendix-otherMoleculesduringSingleFlare}
\end{figure*}

To provide full transparency regarding the behaviour of other prominent species that were not discussed in Sect. \ref{subsec: Results-chemical composition}, we show the evolution of $\COtwo$, $\water$, H, $\OH$, and $\NHtwo$ during the first $\sim$$2.3$ days ($\Pwind/2$) after the flare event in Fig. \ref{fig: appendix-otherMoleculesduringSingleFlare}.

%\newpage
\clearpage
%\nopagebreak

\section{Transmission spectra of \textsc{run 2} (Sect. \ref{subsec: Results-different climates})}
\label{app: Run 2 transmission spectra}

\begin{figure*}[hb!]
	\centering
	\includegraphics[width=17cm]{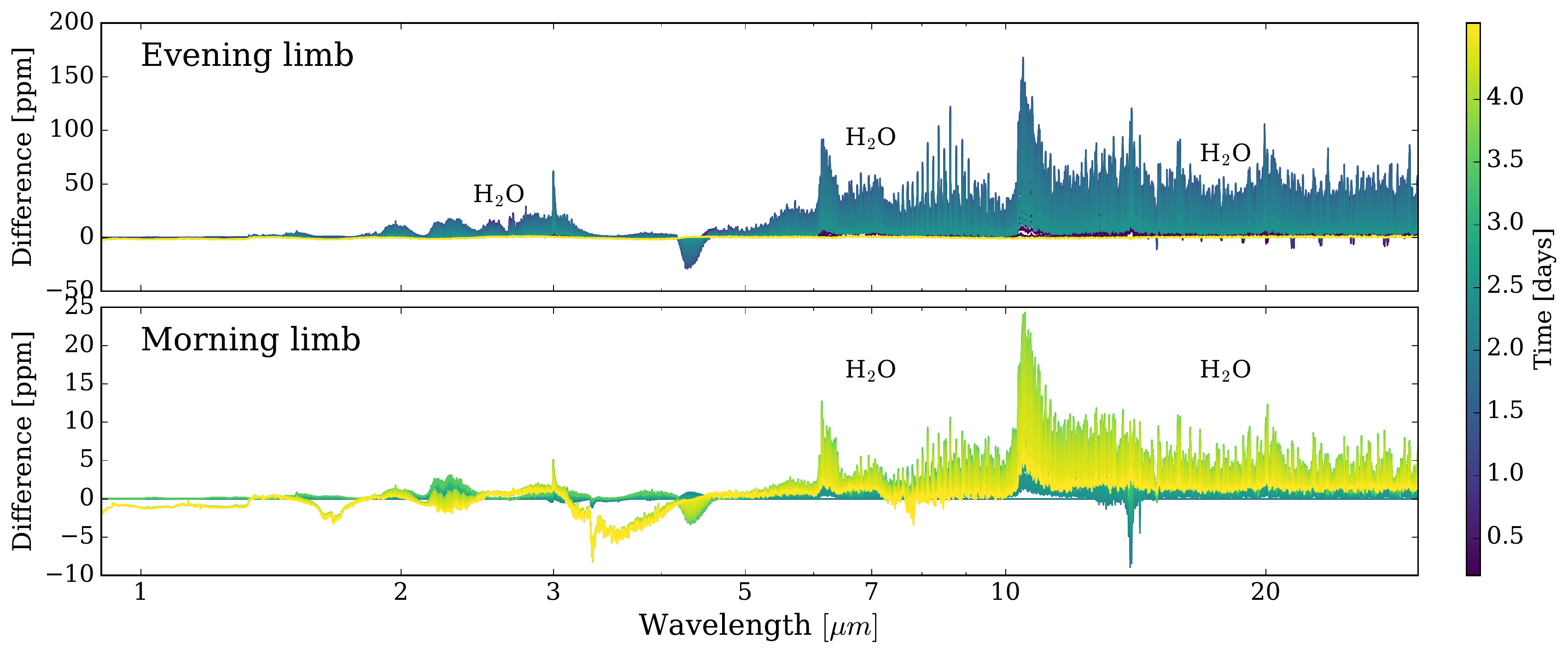}
	\includegraphics[width=17cm]{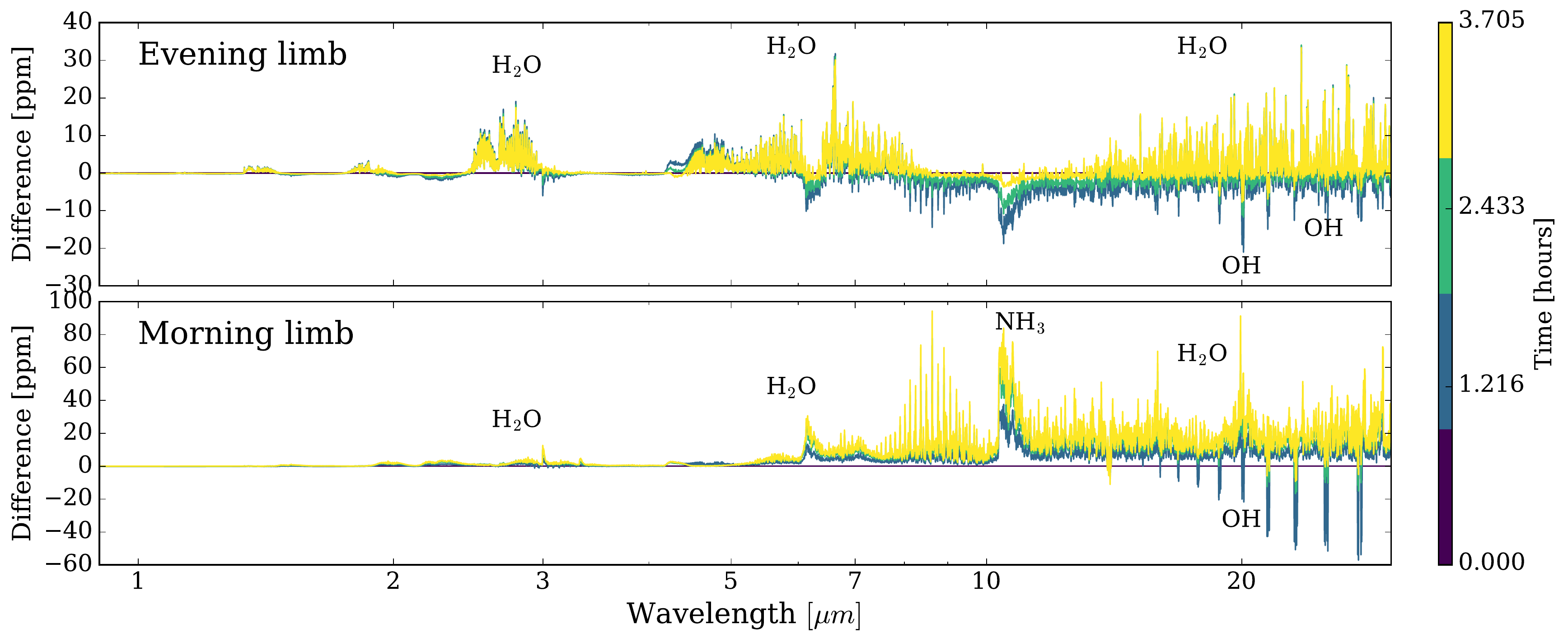}
	\caption{Differences in the transmission spectra of the evening and morning limb with respect to the pre-flare state during the flare (\textit{lower}) and up to $\Pwind$ after the flare (\textit{upper}) for \textsc{run 2} of Sect. \ref{subsec: Results-different climates}.}
	\label{fig: appendix-transSpecRun2-evolutionOfTrSpec}
\end{figure*}

We briefly discuss the evolution of the transmission spectra during and after the flare event of \textsc{run 2} (planet model with $\Teff =1600$ K and $\Pwind =4.6$ days; see Table \ref{tab: Results-Climates-Runs}) based on Fig. \ref{fig: appendix-transSpecRun2-evolutionOfTrSpec}.
In contrast to the flare simulations on the $\Teff=800$ K planet, we see that mainly $\water$ affects the transmission spectrum in the first $\sim$4.6 days after the flare.
Additionally, production of $\OH$ impacts the spectrum above $\sim$$15 \um$.
After $\sim$4.6 days (not shown), the spectrum is altered by less than $\sim$20 ppm, which is in agreement with our finding that the chemical composition returns to its pre-flare values almost completely within the $\Pwind$ time interval.

\clearpage

\section{Evolution of constituents not discussed in Sect. \ref{subsec: Results-ffd}}
\label{app: Other species during the ffd event}

\begin{figure*}
	\centering
	\includegraphics[width=17cm]{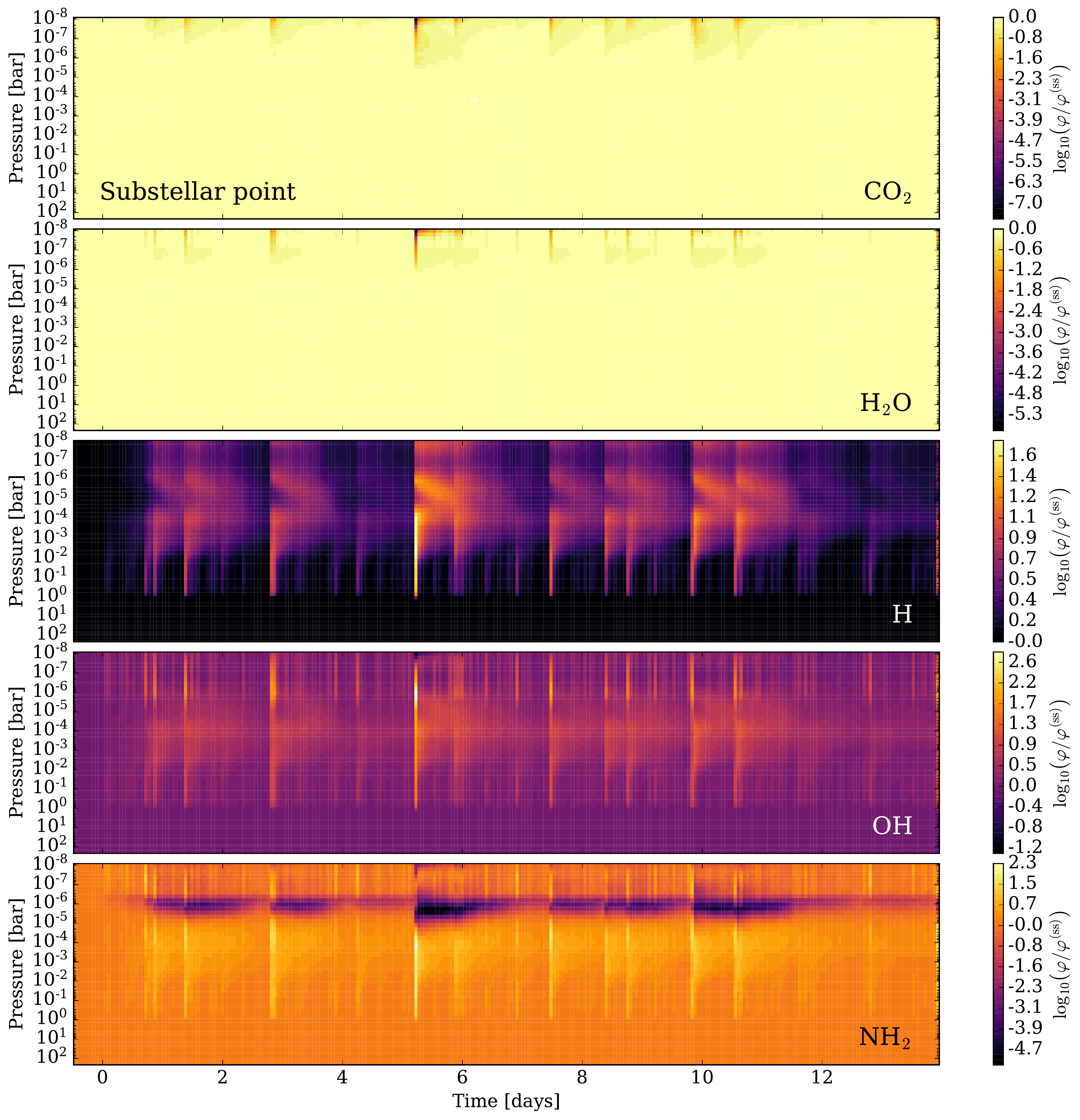}
	\caption{Same as Fig. \ref{fig: results-FFD-2D_plots_sub}, but for $\COtwo$, $\water$, H, $\OH$, and $\NHtwo$.}
	\label{fig: appendix-FFD_substellar}
\end{figure*}

\begin{figure*}
	\centering
	\includegraphics[width=17cm]{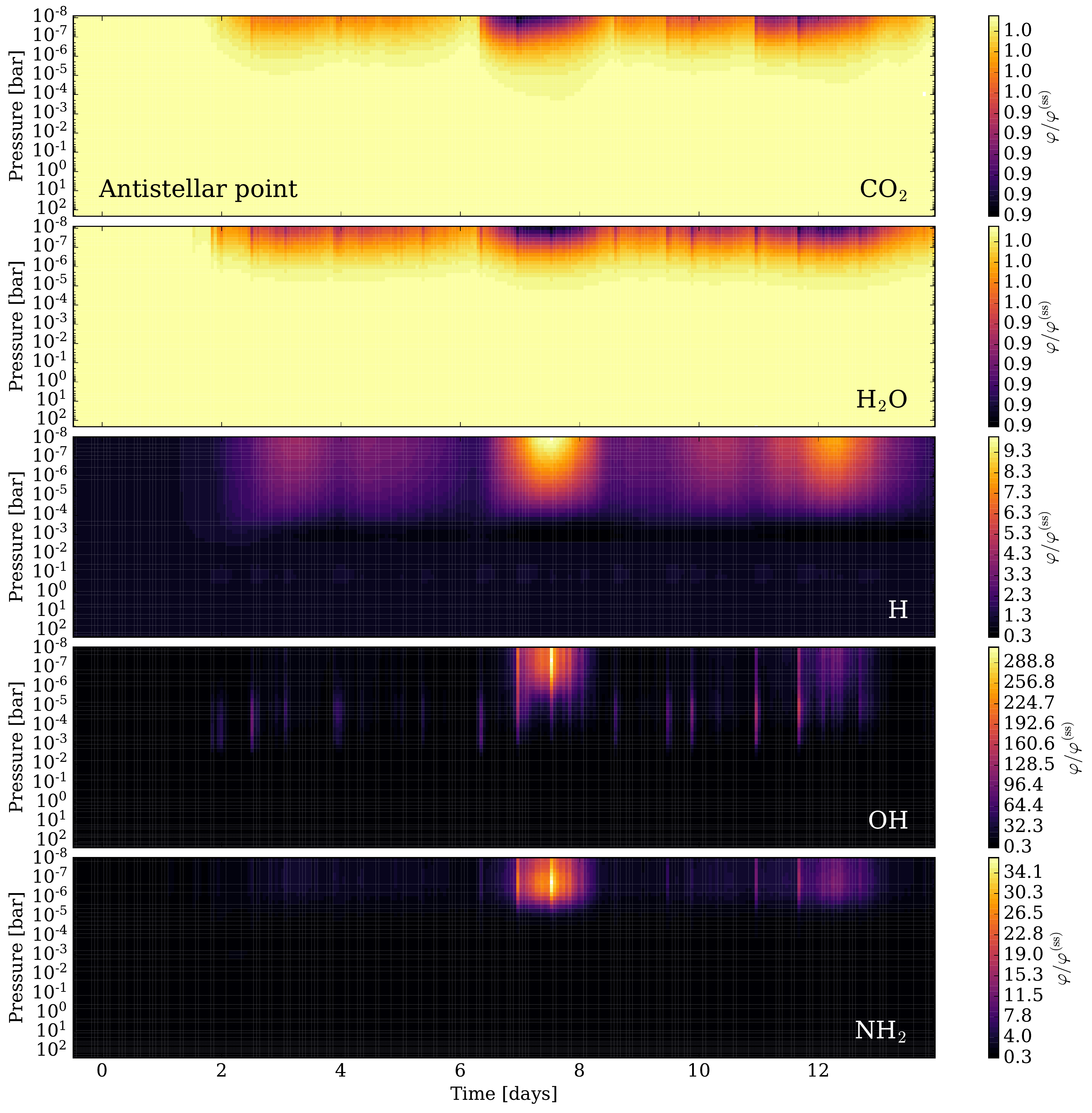}
	\caption{Same as Fig. \ref{fig: results-FFD-2D_plots_anti}, but for $\COtwo$, $\water$, H, $\OH$, and $\NHtwo$.}
	\label{fig: appendix-FFD_antistellar}
\end{figure*}

To provide full transparency regarding the behaviour of other prominent species that were not discussed in Sect. \ref{subsec: Results-chemical composition}, we show the evolution of $\COtwo$, $\water$, H, $\OH$, and $\NHtwo$ throughout the vertical atmosphere on the substellar (Fig. \ref{fig: appendix-FFD_substellar}) and antistellar point (Fig. \ref{fig: appendix-FFD_antistellar}) during $\sim$two weeks of continuous flaring according to a FFD.

\end{appendix}

\end{document}